\documentclass[12pt]{article}

\input epsf.sty

\newcommand{\insertfig}[2]{\mbox{\epsfxsize=#1cm \epsfbox{#2.eps}}}

\font\cmss=cmss12 
\def\1{\hbox{{1}\kern-.25em\hbox{l}}}
\def\bfZ{\relax{\hbox{\cmss Z\kern-.4em Z}}}

\begin{document}
\begin{titlepage}

\begin{flushright}
DOE/ER/40762-008 \\ [-2mm]
UMD-PP\#02-010 \\ [-2mm]
YITP-SB-01-44
\end{flushright}

\centerline{\large \bf Nonlinear evolution in high-density QCD}

\vspace{15mm}

\centerline{\bf I.I. Balitsky$^{a,b}$, A.V. Belitsky$^{c,d}$}

\vspace{15mm}

\centerline{\it $^a$Physics Department, Old Dominion University}
\centerline{\it VA 23529, Norfolk, USA}

\vspace{5mm}

\centerline{\it $^b$Theory Group, Jefferson Laboratory,}
\centerline{\it  VA 23606, Newport News, USA}

\vspace{5mm}

\centerline{\it $^c$C.N.\ Yang Institute for Theoretical Physics}
\centerline{\it State University of New York at Stony Brook}
\centerline{\it NY 11794-3840, Stony Brook, USA}

\vspace{5mm}

\centerline{\it $^d$Department of Physics}
\centerline{\it University of Maryland at College Park}
\centerline{\it MD 20742-4111, College Park, USA}

\vspace{15mm}

\centerline{\bf Abstract}

\vspace{0.5cm}

\noindent
We consider deeply inelastic scattering at very high energies in the saturation
regime. The emerging picture corresponds to the propagation of a dipole, the
quark-antiquark pair, in a shock wave color field of the target. We use the 
fomalism of Wilson lines to study the evolution of dipole densities in energy 
logarithms. Our analysis results into an equation in multicolor limit which sums 
leading logs but keeps the nonlinearities up to cubic order in densities.

\vspace{1.5cm}



\end{titlepage}

\section{QCD of dense parton systems.}

A standard tool in QCD analyses of the deeply inelastic lepton-nucleon,
$\ell N \to \ell' X$, scattering (DIS) for moderate values of Bjorken
variable, $x_{\rm B}$, is the factorization theorems and
Dokshitzer-Gribov-Lipatov-Altarelli-Parisi (DGLAP) evolution equation
\cite{DGLAP}. This equation describes the change of the DIS
observables with the change of a resolution scale, which for the
process in question is the virtuality of the probe $Q^2$. The latter
serves as a microscope which allows to penetrate deep inside the hadron
substructure and observe its constituents with transverse size
$\delta x_\perp \sim Q^{-1}$ and longitudinal extent $\frac{1}{x_{\rm B}}$.

The DGLAP dynamics is based on the separation of the DIS amplitude into a
`hard' part coming from the transverse momenta $k_{\perp}^2 > \mu^2$ and
a `soft' part coming from low $k_{\perp}^2 < \mu^2$, where $\mu^2$ is a
scale dividing short and long distance physics. The incoherence of these
phenomena allows for a factorization of infrared part into a universal
matrix element of a non-local composite light-cone operators constructed
from quark and gluon fields. On the other hand, the contribution from
hard momenta gives the coefficient functions. The factorization scale $\mu^2$
serves as a normalization point for those operators. The change in $\mu^2$
is governed by the conventional renormalization group equations. Taking
$\mu^2 = Q^2$, we come to the usual result that the $Q^2$ dynamics of DIS
cross sections is driven by the renormalization group equations for the
light-cone operators. In terms of QCD perturbation theory it results from
the summation of the contributions of the type $\left( \alpha_s \ln Q^2
\right)^n$, etc., in momentum transfer.

A very important property of the factorization alluded to above is that the
coefficient functions are purely perturbative. Indeed, the effective
coupling constant is determined by characteristic transverse momenta so that
the contributions coming from large $k_{\perp}^2 > \mu^2$ are treatable
within QCD perturbation theory as long as $\mu^2$ is sufficiently large. The
nonperturbative physics enters the game only when we lower the normalization
point $\mu^2$ down to a typical hadronic scale of order $\sim 1$ GeV. The
higher order terms of perturbative expansion, for both the coefficient
functions and the anomalous dimensions of the light-cone operators, lie in
the same framework of linear evolution and lead to corrections $\sim\alpha_s$,
$\alpha_s^2$, etc. Thus, to compare experimental measurements
of structure functions ${\cal F} (x_{\rm B}, Q^2)$ at different $Q^2$ we rely
only on perturbative QCD and the linear character of the DGLAP equations
makes this comparison  especially simple.

The situation changes drastically if one is interested in the domain of small
$x_{\rm B}$. The DGLAP evolution leads to a strong rise of the DIS structure
function
\begin{eqnarray}
x_{\rm B} {\cal F} (x_{\rm B}, Q^2)
\sim \exp \left( {\rm const.} \ln \frac{1}{x_{\rm B}} \ln\ln Q^2 \right)^{1/2}
\, ,
\end{eqnarray}
at small values of Bjorken variable $x_{\rm B}$. If one bears on using the
DGLAP evolution for smaller and smaller $x_{\rm B}$, higher loop contributions
become enhanced by additional factors $\ln \frac{1}{x_{\rm B}}$ and the
perturbative expansion of the coefficient functions and anomalous dimensions
breaks down calling  for the small-$x_{\rm B}$ resummation. Recall that DGLAP
equation sums logs of the hard scale to all orders, i.e., terms of the kind
$\left( \alpha_s \ln Q^2 \right)^n$ and $\left( \alpha_s \ln Q^2 \ln
\frac{1}{x_{\rm B}} \right)^n$. It takes only a single logarithm or none
of the energy for a power of the coupling constant. Thus, it fails for very
low-$x_{\rm B}$ when $\alpha_s \ln \frac{1}{x_{\rm B}} \gg 1$ and these
contributions have to be summed over. In perturbative QCD, the small-$x_{\rm B}$
asymptotic behaviour is described in the leading logarithmic approximation (LLA)
by the Balitsky-Fadin-Kuraev-Lipatov (BFKL) pomeron \cite{BFKL} which sums up
the leading energy logarithms $\left( \alpha_s \ln \frac{1}{x_{\rm B}} \right)^n$.

Unfortunately, the BFKL evolution, for a review, see \cite{Lip97}, suffers
from its own caveats. The first one is the lack of unitarity: the power
behavior of the cross section due to BFKL dynamics
\begin{eqnarray}
\label{BFKLrise}
x_{\rm B} {\cal F} (x_B, Q^2)
\sim x_{\rm B}^{\alpha_{\rm I\!P}} ,
\qquad\qquad
\alpha_{\rm I\!P} = 1 + 4 N_c \frac{\alpha_s}{\pi} \ln 2
\, ,
\end{eqnarray}
violates the the so-called Froissart theorem stating that a cross section may
grow at most as $\ln^2 \frac{1}{x_{\rm B}}$ at $x_{\rm B} \to 0$. Obviously,
this result means that approximations involved in the derivation of the
BFKL equation become inadequate and in order get the true asymptotic behaviour
at small $x_{\rm B}$, we must go beyond the LLA. Unlike the DGLAP case, this
is not a purely technical problem of calculating loop corrections to the
kernels. There are $\alpha_s$ corrections to the BFKL kernel \cite{FadLip98},
but in addition there are unitarity corrections which go beyond the framework
of the BFKL equation. At small $\alpha_s$ and $x_{\rm B}$, the latter
corrections seem to dominate over the next-to-leading BFKL effects \cite{Run98}.

The second problem with the BFKL evolution is its infrared instability. We
can safely apply perturbative QCD to the small-$x_{\rm B}$ DIS if the
characteristic transverse momenta of the gluons $k_{\perp}$ in the gluon
ladder are large. For the first few evolution steps, one can check by an
explicit calculation that the characteristic $k^2_{\perp}$ are of the order
$\sim Q^2$. However, as $x_{\rm B}$ decreases, it turns out that the
characteristic transverse momenta in the middle of the gluon ladder drift
towards $\Lambda_{\rm QCD}$ making the application of perturbative QCD
questionable. This is related to the fact that the operator expansion for
the high-energy scattering in terms of Wilson line operators, which represent
quarks moving with almost the velocity of light, \cite{Bal96} is based on the
factorization in the rapidity, $\eta \equiv \ln \frac{1}{x_{\rm B}}$,
\cite{Ste95,Bal98} rather than the transverse momentum. Unlike the usual
light-cone expansion, the high-energy expansion in Wilson operators does not
admits an additional meaning of perturbative versus nonperturbative separation.
Contrary, both the coefficient functions and the matrix elements have
perturbative as well as nonperturbative parts. This happens because, as we
mentioned above, the coupling constant in a scattering process is determined by
the scale of the transverse momenta. When we use the factorization in hard
($k_{\perp} > \mu$) and soft ($k_{\perp} < \mu$) momenta, we calculate the
coefficient functions perturbatively, since $\alpha_s(k_{\perp} > \mu)$ is
small, whereas the matrix elements are nonperturbative. Conversely, when we
factorize the amplitude in rapidity, both fast and slow parts have contributions
coming from regions of large and small $k_{\perp}$.  In this sense, the
small-$x_{\rm B}$ evolution in QCD is not protected from the infrared side in
the same way as the DGLAP evolution is: in order to compare the two structure
functions measured at different (small) values of $x_{\rm B}$ the perturbative
QCD may be insufficient and, in order to explain the small-$x_{\rm B}$ behavior
of structure functions, it may be necessary to take into account the interplay
between the hard and soft pomeron.

Both of these problems can be resolved simulteneously if, as argued in
\cite{GriLevRys83,MueQiu86,MclVen94,Mue99a} the partons in the highly
energetic nucleon reach the state of saturation: the recombination of
partons balances the rise of the cross section due to parton emission, and
the hard saturation scale $Q_s$ sets the scale of the effective coupling
constant. Indeed, once we have a rather dense gluon system created by
conventional parton splitting described by the linear DGLAP and BFKL
evolution, the partons populating a given space-time volume inside the
hadron start to overlap and an absorption competes with creation. This is
expected to happen when gluons occupy the entire transverse area of the
hadron disc ${\mit\Sigma}^\perp_{\rm hadr} = \pi R^2$,
i.e.,
\begin{eqnarray*}
\frac{{\mit\Sigma}^\perp_{\rm hadr}}{{\mit\Sigma}^\perp_{\rm part}} \sim
\frac{\pi R^2}{ \left( \delta x_\perp \right)^2 n}
\sim 1 \, ,
\end{eqnarray*}
where the number of partons $n$ is proportional to the gluon density
$x_{\rm B} G (x_{\rm B}, Q^2)$ which overwhelms quarks at small $x_{\rm B}$.
This is a result of parton saturation which is expected to tame the growth
of their number. So in the case of gluon overlap, they start to interact
strongly although the QCD coupling may well still be in the perturbative
domain. This regime was addressed in the pioneering work by Gribov, Levin and
Ryskin \cite{GriLevRys83} and has resulted into a suggestion of a first
nonlinear evolution equation which goes under their names, the GLR equation.
It received an early discussion in Ref.\ \cite{Snow84}. A derivation of
the later within double logarithmic approximation has been given in
\cite{MueQiu86}. The question of the value of the scale at which the
annihilation takes over the production based on the analysis and solutions
\cite{GriLevRys83,ColKwi90,BarSchBlu91,BarLev92} to the GLR equation
has resulted into the aforementioned concept of saturation.

\begin{figure}[t]
\begin{center}
\hspace{0cm} \mbox{
\begin{picture}(0,100)(100,0)
\put(10,0){\insertfig{6}{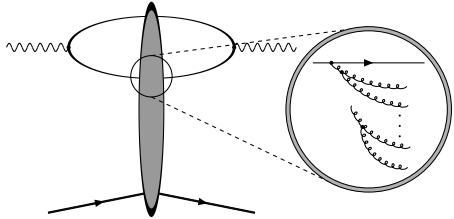}}
\end{picture}
}
\end{center}
\caption{\label{dis} Scattering of the photon probe off the hadron at
high energies. The virtual quantum fluctuates into a bunch of dipoles
which interact with the target.}
\end{figure}

A new era of investigations of nonlinear QCD phenomena has been been
initiated by the experimental results from HERA on small-$x_{\rm B}$
measurement of the DIS structure functions. It created diverse
dynamical approaches to the corresponding physics which we will
discuss below and has led a phenomenologically successful saturation
models, see e.g., \cite{GolWus98}.

The most transparent picture of the underlying nonlinear effects arises
in the dipole frame \cite{MueSal96,Mue99}, where the hadron carries almost
the entire rapidity, and moves with almost the speed of light, but still the
virtual photon is quite energetic. The incoming photon fluctuates into
quark-antiquark pair, a dipole, and interacts with the target via
gluon emission. Since the hadron is contracted due to Lorentz dilation,
the dipole with accompanying radiation sees it as a color source of a
transverse extent living on the light-cone, see Fig.\ \ref{dis}.
This color field is created by the constituents of the well developed wave
function of the hadron which in view of its high intensity, i.e., big
occupation numbers, can be considered as classical. The strength of the
field in the regime of saturation is $1/\sqrt{\alpha_s}$. This can be
achieved either by going to the very low $x_{\rm B}$ for a hadron or
scattering off a nucleus \cite{MclVen94}, or both. Actually in the present
paper we will consider the last possibility which justifies the neglect
of the so-called Pomeron loops which are suppressed then by the atomic
number of the nucleus as compared to exchange contributions. At high energies,
we can neglect the recoil of emitted gluons so that it is legitimate to use
the eikonal approximation \cite{Nac91,ColEll91}. The proper degrees of freedom
for the fast particles moving along the straight trajectories collinear to
their velocities are the infinite ordered gauge factors, mentioned earlier
with respect to an operator approach to the high energy scattering. In this
case, the color dipole is given as two Wilson lines stretched along the
light-like direction $n_\mu = \frac{1}{\sqrt{2}} (1, \mbox{\boldmath$0$}, 1)$,
\begin{eqnarray}
\label{PathExp}
{\cal U} \left( \infty, - \infty; \mbox{\boldmath$x$} \right)
\equiv P \exp \left(
i g \int_{- \infty}^{\infty} dx_- A_+ (x_-, \mbox{\boldmath$x$})
\right) \, ,
\end{eqnarray}
and separated by the transverse distance $x_{\perp\mu} - y_{\perp\mu}
= (0, \mbox{\boldmath$x$} - \mbox{\boldmath$y$}, 0)$, i.e.,
\begin{equation}
{\cal N} (\mbox{\boldmath$x$}, \mbox{\boldmath$y$})
= \langle 0 | T \, \frac{1}{N_c} {\rm tr}
\left(
{\cal U} \left( \infty, - \infty; \mbox{\boldmath$x$} \right)
{\cal U}^\dagger \left( \infty, - \infty; \mbox{\boldmath$y$} \right)
- \1
\right)
| 0 \rangle \, .
\end{equation}
It is evaluated in the external field of the hadron or nucleus alluded to
above. The structure functions are given by a convolution of the probability
for the photon to fluctuate into the quark-antiquark pair and the dipole
cross section expressed as an integral of the dipole density over the impact
parameter, $\mbox{\boldmath$b$} \equiv \frac{1}{2} \left( \mbox{\boldmath$x$}
+ \mbox{\boldmath$y$} \right)$. Namely, \cite{NikZak90},
\begin{equation}
{\cal F} (x_{\rm B}, Q^2) \sim
\int d^2 \mbox{\boldmath$z$} \int d z \,
| {\mit\Psi}^\gamma_{\bar q q} (z, \mbox{\boldmath$z$}, Q^2) |^2
\int d^2 \mbox{\boldmath$b$} \,
{\cal N} (\mbox{\boldmath$x$}, \mbox{\boldmath$y$})
\, ,
\end{equation}
with $\mbox{\boldmath$z$} \equiv \mbox{\boldmath$x$} - \mbox{\boldmath$y$}$.

In doing this procedure one introduces a separation in rapidity as discussed
previously, according to which the gluons of high rapidity go to the impact
factor, --- the square of the photon wave function alluded to above, --- while
the slow fields form the dipole. The change of this divide cannot change a
physical observable and it is governed by an evolution equation for ${\cal N}$.
To find the dipole density at higher rapidity $\eta + \delta \eta$, i.e.,
smaller Bjorken variables, in terms of the one at $\eta$ we have to integrate
the fast gluon modes out in the strip $\delta \eta$. At low densities, this
evolution stems from an independent branching of dipoles and results into the
linear BFKL equation \cite{Mue94}. Saturation effects manifest themselves in
breaking of this pattern when the dipoles start to `feel' each other and
intensively interact. Thus, there are two types of corrections to this result:
First, radiative corrections to the photon wave functions, and second,
contributions of higher, or multiple, dipole densities to the cross section. The
latter arise as soon as an additional gluon is emitted along from one of the
quarks in the pair. In the multicolor limit this quark-gluon-antiquark system is
reduced to the pair of dipole densities both of which interact with the target
field. $n$ extra gluons lead to $n + 1$ dipoles, etc., which are accompanied in
the structure function ${\cal F}$ by the multi-parton photon wave functions,
i.e., schematically, $|{\mit\Psi}^\gamma_{\bar q (n g) q}|^2 {\cal N}^{n + 1}$.

A number of studies along this line has led to a nonlinear equation which
generalized GLR equation. It was derived first in \cite{Bal96} where an
infinite set of coupled integral-differential equations has been given and
a perturbative kernel for the first nonlinearity has been found. In Ref.\
\cite{Kov99} the latter was deduced from the Mueller's nonlinear equation
for the generating functional of dipole densities \cite{Mue94} in the
multicolor limit. In \cite{Bra00} it was rederived using the direct summation
of the fan diagrams. In \cite{IanLeoMcl01b,IanLeoMcl01a} it was deduced from the
functional equation, for the statistical weights of the Color Glass Condensate,
a state of dense gluon matter at high energies for heavy nucleus
\cite{IanLeoMcl01a,McL01}, derived in \cite{JalKovLeoWei97} and rederived in
\cite{IanLeoMcl01b,IanLeoMcl01a}. The paper \cite{Wei00} dealt with the infinite
hierarchy of equations alluded to above which were summarized in a compact form
of the functional Fokker-Plank equation and its properties has been discussed.

The most general nonlinear evolution equation in the large-$N_c$ limit reads
\begin{eqnarray}
\label{GeneralNonlinear}
\frac{d}{d \ln \frac{1}{x_{\rm B}}}
{\cal N} (\mbox{\boldmath$x$}, \mbox{\boldmath$y$})
\!\!\!&=&\!\!\!
\int d^2 \mbox{\boldmath$z$} \
{\cal K}_1 (\mbox{\boldmath$x$}, \mbox{\boldmath$z$}, \mbox{\boldmath$y$})
\
\Big\{
{\cal N} (\mbox{\boldmath$x$}, \mbox{\boldmath$z$})
+
{\cal N} (\mbox{\boldmath$z$}, \mbox{\boldmath$y$})
\Big\}
+
\int d^2 \mbox{\boldmath$z$} \
{\cal K}_2 (\mbox{\boldmath$x$}, \mbox{\boldmath$z$}, \mbox{\boldmath$y$})
\  {\cal N} (\mbox{\boldmath$x$}, \mbox{\boldmath$z$})
{\cal N} (\mbox{\boldmath$z$}, \mbox{\boldmath$y$})
\nonumber\\
&+&\!\!\!
\int d^2 \mbox{\boldmath$z$} \ d^2 \mbox{\boldmath$z$}' \
{\cal K}_3 (\mbox{\boldmath$x$}, \mbox{\boldmath$z$},
\mbox{\boldmath$z$}', \mbox{\boldmath$y$})
\
{\cal N} (\mbox{\boldmath$x$}, \mbox{\boldmath$z$})
{\cal N} (\mbox{\boldmath$z$}, \mbox{\boldmath$z$}')
{\cal N} (\mbox{\boldmath$z$}', \mbox{\boldmath$y$})
+ \dots
\, .
\end{eqnarray}
The necessity to take the multicolor limit will be explained in the main
text. Here it is sufficient to say that it allows the factorization of
multiple Wilson line correlations into dipole densities and leads to a closed
equation (\ref{GeneralNonlinear}). Note that the term ${\cal K}_n$ starts
from the $\alpha_s^{n - 1}$ order in perturbation theory. The known evolution
kernels are ${\cal K}_1$ and ${\cal K}_2$. The former, obviously, coincides
with the BFKL evolution kernel and is given by
\begin{eqnarray}
\label{BFKLkernel}
&&{\cal K}_1 (\mbox{\boldmath$x$}, \mbox{\boldmath$z$}, \mbox{\boldmath$y$})
= \frac{\alpha_s}{2\pi^2} N_c
\Bigg\{
\frac{( \mbox{\boldmath$x$} - \mbox{\boldmath$y$} )^2 }
{( \mbox{\boldmath$z$} - \mbox{\boldmath$x$} )^2
( \mbox{\boldmath$z$} - \mbox{\boldmath$y$} )^2}
\\
&&\qquad\qquad\qquad\qquad\qquad
- \frac{1}{2}
\left(
\delta^{(2)} (\mbox{\boldmath$z$} - \mbox{\boldmath$x$})
+
\delta^{(2)} (\mbox{\boldmath$z$} - \mbox{\boldmath$y$})
\right)
\int d^2 \mbox{\boldmath$z$}'
\frac{( \mbox{\boldmath$x$} - \mbox{\boldmath$y$} )^2 }
{( \mbox{\boldmath$z$}' - \mbox{\boldmath$x$} )^2
( \mbox{\boldmath$z$}' - \mbox{\boldmath$y$} )^2}
\Bigg\}
+ {\cal O} \left( \alpha_s^2 \right)
\, .
\nonumber
\end{eqnarray}
Use also ${\cal N} (\mbox{\boldmath$x$}, \mbox{\boldmath$x$}) = 0$.
The kernel for the first nonlinearity has been found in a number of
studies to be
\begin{equation}
\label{LOnonLkernel}
{\cal K}_2 (\mbox{\boldmath$x$}, \mbox{\boldmath$z$}, \mbox{\boldmath$y$})
= \frac{\alpha_s}{2\pi^2} N_c
\frac{( \mbox{\boldmath$x$} - \mbox{\boldmath$y$} )^2 }
{( \mbox{\boldmath$z$} - \mbox{\boldmath$x$} )^2
( \mbox{\boldmath$z$} - \mbox{\boldmath$y$} )^2}
+ {\cal O} \left( \alpha_s^2 \right) \, .
\end{equation}

Extensive numerical and semi-analytical solutions to the generalized
nonlinear equation with ${\cal K}_{n > 2} = 0$ performed in
\cite{AyaGayLev97,Bra00,LevTuc00,KimKwiMar01,LevLub00,IanMcl01} have
demonstrated the desired suppression of the growth in the parton densities
and resulted into quantitative estimates of the saturation scale $Q_s$
where this turnover actually takes place. In Ref.\ \cite{Bal01} the
solution to the equation has been expressed in terms of a path integral
suitable for lattice evaluations.

Our current study is devoted to the computation of the evolution kernel
${\cal K}_3$. Presently, we neglect the next-to-leading BFKL terms in ${\cal K}_1$
and ${\cal O} (\alpha_s)$ corrections to the three-pomeron vertex ${\cal K}_2$
and reserve their evaluation for our future study. However, one can expect them
to be parametrically less important that ${\cal K}_3$ since a bulk of these
${\cal O} (\alpha_s)$ corrections comes from the effects of running coupling
constant. For rather small coupling constant, the unitarity effects, which cure
the strong rise (\ref{BFKLrise}), become dominant at rapidities \cite{Mue96}
\begin{eqnarray*}
\eta \sim \frac{2}{\alpha_{\rm I\!P} - 1} \ln \frac{1}{\alpha_s} \, ,
\end{eqnarray*}
which are smaller than the rapidities when the running of the QCD coupling in the
BFKL equation starts to be important. The latter was estimated to be \cite{Run98}
\begin{eqnarray*}
\eta \sim \alpha_s^{-5/3} \, .
\end{eqnarray*}
Therefore, the unitarity restoring effect dominate long before the diffusion
constraints, and the question of unitarity restoration can be considered
within the fixed coupling approximation \cite{Mue96}.

Those readers who are not interested in technical details of the derivation,
leading to the final result, can skip sections the most of the consequent
presentation and go to section \ref{EvolutionKernelK3} and conclusions. For
the rest, our paper is organized as follows. In the next section we consider
the Regge limit of the deeply inelastic scattering amplitude and define the
gluon field configuration of the boosted hadron which serves as a scattering
source for the virtual photon. Then, in section \ref{PropagatorsExtField} we
compute the exact quark and gluon propagators in this shock-wave field. We
use them in section \ref{LOandNLOnonlinearities} for a calculation of leading
and next-to-leading nonlinearities in the generalized evolution equation.
Finally, we present a few concluding remarks. The two technical appendices
give a detailed account of two-loop computation of diagrams with nonabelian
vertices and two-dimensional Fourier transformation from the momentum to the
coordinate space.

\section{High-energy limit and shock-wave.}

As we already explained in detail earlier in the dipole frame the
external electromagnetic probe, the photon, having momentum $q$
fluctuates into a quark-antiquark pair and passes through the Lorentz
contracted color field of the hadron or nucleus with momentum $p$.
Let us determine this field configuration in the high-energy limit. To
this end we consider the path integral which defines the DIS amplitude
\begin{eqnarray}
\label{def-HadTen}
T_{\mu\nu} (q, p)
\!\!\!&=&\!\!\!
i \int d^4 z \, {\rm e}^{i z \cdot q}
\langle p | T j_\mu (z/2) j_\nu (-z/2) | p \rangle_A
\nonumber\\
\!\!\!&\equiv&\!\!\!
i \int d^4 z {\rm e}^{i z \cdot q}
\langle p |
\int D A \, D \psi \, D \bar\psi \,
\exp \left( i \int d^4 y \, {\cal L} (y) \right)
j_\mu (z/2) j_\nu (-z/2)
| p \rangle
\, .
\end{eqnarray}
Each field configuration is weighted with the exponential of the QCD
Lagrangian ${\cal L} = {\cal L}_{\rm cl} + {\cal L}_{\rm gf}$. Here the
classical part is
\begin{eqnarray}
\label{YM-renormalized}
{\cal L}_{\rm cl} &=& - \frac{1}{4} \left(G^a_{\mu\nu}\right)^2
+ \bar\psi i \!\not\!\! {\cal D} \psi \, ,
\end{eqnarray}
and the gauge fixing part will be specified later in this section complying
with the form of the external color source which will make the treatment
especially simple in the present circumstances. Our conventions are fixed by
the following definitions for the covariant derivative ${\cal D}_\mu =
\partial_\mu - i g A_\mu$ and the field strength tensor $G^a_{\mu \nu} =
\partial_\mu A^a_\nu - \partial_\nu A^a_\mu + g f^{abc} A^b_{\mu} A^c_{\nu}$.

Motivated by the demand outlined in the introduction we introduce two
light-like vectors $n_\mu$ and $n^\star_\mu$ such that $n^2 = n^{\star\, 2} = 0$,
and $n \cdot n^\star = 1$. We use the conventions for light-cone coordinates
$z_+ \equiv z \cdot n$, $z_- \equiv z \cdot n^\star$, so that $z_\mu = z_- n_\mu
+ z_+ n^\star_\mu + z_{\perp\mu}$. Thus $z^2 = 2 z_+ z_- - \mbox{\boldmath$z$}{}^2$,
where $z_\perp^2 = - \mbox{\boldmath$z$}{}^2$. The projection on the transverse
plane is done with the transverse metric $g^\perp_{\mu \nu} = g_{\mu \nu}
- n_\mu n^\star_\nu - n_\nu n^\star_\mu$.

In massless limit the target four-momentum is light-like $p = n^\star$ and
the virtual photon vector can be decomposed in Sudakov variables as follows
\begin{eqnarray*}
q_\mu = - x_{\rm B} n^\star_\mu - \frac{q^2}{2 x_{\rm B}} n_\mu \, ,
\end{eqnarray*}
with $x_{\rm B} \equiv - q^2/(2 q \cdot p)$. The small-$x_{\rm B}$ limit,
$\lambda \equiv 1/x_{\rm B} \to \infty$, can be treated as a rescaling of
the fields in the functional integral
\cite{VerVer93}
\begin{eqnarray}
\lim_{x_{\rm B} \to 0} T_{\mu\nu} (q, p)
\!\!\!&=&\!\!\!
i \int d^4 z {\rm e}^{- i z_- - i z_+ q^2/2} \nonumber\\
&&\times\lim_{\lambda \to \infty}
\langle p | T
j_\mu \left(
\lambda^{-1} {z_+}/{2},
\lambda {z_-}/{2},
{\mbox{\boldmath$z$}}/{2}
\right)
j_\nu
\left(
-  \lambda^{-1} {z_+}/{2},
- \lambda {z_-}/{2},
- {\mbox{\boldmath$z$}}/{2}
\right)
| p \rangle \, .
\end{eqnarray}
After changing the variables in the path integral we have the amplitude
\begin{eqnarray}
\lim_{x_{\rm B} \to 0} T_{\mu\nu} (q, p) =
i \int d^4 z {\rm e}^{- i z_- - i z_+ q^2/2}
\langle p | T j_\mu (z/2) j_\nu (-z/2) | p \rangle_B \, .
\end{eqnarray}
evaluated in the background gluon field
\begin{eqnarray}
\label{RescaledField}
B_+ (z_+, z_-, \mbox{\boldmath$z$})
\!\!\!&=&\!\!\!
\lim_{\lambda \to \infty}
\lambda A_+
\left( \frac{z_+}{\lambda}, \lambda z_-, \mbox{\boldmath$z$} \right)
\, , \nonumber\\
B_- (z_+, z_-, \mbox{\boldmath$z$})
\!\!\!&=&\!\!\!
\lim_{\lambda \to \infty}
\frac{1}{\lambda} A_+
\left( \frac{z_+}{\lambda}, \lambda z_-, \mbox{\boldmath$z$} \right)
\, , \nonumber\\
\mbox{\boldmath$B$} (z_+, z_-, \mbox{\boldmath$z$})
\!\!\!&=&\!\!\!
\lim_{\lambda \to \infty}
\mbox{\boldmath$A$}
\left( \frac{z_+}{\lambda}, \lambda z_-, \mbox{\boldmath$z$} \right)
\, .
\end{eqnarray}
Taking the limit $\lambda \to \infty$ one gets from Eq.\ (\ref{RescaledField})
\begin{equation}
\label{ExternalField}
B_+ = \delta (z_-) \beta (\mbox{\boldmath$z$}),
\qquad
B_- = \mbox{\boldmath$B$} = 0 \, .
\end{equation}
Thus, the background field has the form of a shock-wave type. The field
strength tensor for this potential
\begin{equation}
G_{\mu \nu} = \delta (z_-)
\left(
\partial^\perp_\mu n^\star_\nu
-
\partial^\perp_\nu n^\star_\mu
\right)
\beta (\mbox{\boldmath$z$}) \, ,
\end{equation}
is also localized on the light cone.

Thus, considering the scattering of a dipole in the external shock-wave
field we decompose the total gauge field into its background and quantum
parts
\begin{equation}
\label{ClasQuantDecomp}
A = B + b \, .
\end{equation}
While developing the perturbation theory with respect to the quantum fields
we will keep classical source effects exactly. To make sense out of the
path integral we have to impose a gauge condition on the quantum field $b$.
As it is almost obvious from these considerations, the light-like gauge
$b_- \equiv b \cdot n^\star = 0$ is the most appropriate choice. For this
case, the addendum ${\cal L}_{\rm gf}$ reads
\begin{equation}
{\cal L}_{\rm gf} = - \frac{1}{2 \xi} \left( b^a_- \right)^2
+ \bar \omega^a {\cal D}^{ab}_- \omega^b \, .
\end{equation}
Since the DIS amplitude is gauge independent we take the limit $\xi \to 0$ in
the generating functional (\ref{def-HadTen}) which simplifies considerably the
gluon propagator.

The procedure we have just performed can be translated into the language
of the Wilson renormalization group. We have included into the classical
field the low frequency excitations of the gauge fields $p_{0 -} <
{\mit\Lambda}_0 \sim 1/x_{{\rm B} 0}$, while the quantum part involves
fast configuration. Since the generating functional does not depend on
the separation scale ${\mit\Lambda}_0$ the change in the latter is
compensated by the renormalization group flow, i.e., in each evolution
step we integrate out the quantum field with the momenta in the strip
\begin{equation}
\label{Strip}
p_{0-} < p_- < p_{0-} + \delta p_- \, ,
\end{equation}
in order to go from the scale $p_{0-} + \delta p_-$ to $p_{0-}$.

\section{Propagators in the shock-wave background.}
\label{PropagatorsExtField}

As a next natural step we have to find particle propagators in the
shock-wave background (\ref{ExternalField}). We do this by an explicit
summation of the gluon emission diagrams. For other related discussions
of the backround gauge propagators, the reader is referred to Refs.\
\cite{MclVen95,Bal96,HebWei98,KovMilWei00,IanLeoMcl01a}. Since we will
use both momentum and coordinate space representations we define the
Fourier transformed propagator according to
\begin{equation}
G (p, p')
\equiv
\int d^4 z \, d^4 z' \, {\rm e}^{i p \cdot z - i p' \cdot z'}
G (z, z') \, .
\end{equation}
And the Fourier transform of the shock-wave field is
\begin{equation}
\label{Shock-momentum}
B^a_\mu (k)
= n^\star_\mu \, 2\pi \delta (k_-) \beta^a (\mbox{\boldmath$k$}) \, .
\end{equation}
Now, we address the quark and gluon Green functions in turn.

\begin{itemize}
\item Quark propagator: $i S^i{}_j (x, y) =
\langle 0 | T  \psi^i (x) \bar\psi_j (y) | 0 \rangle_{B}$.
\end{itemize}
Summing up the gluon emission diagrams with the field (\ref{Shock-momentum})
we find the following matrix representation for the propagator
\begin{equation}
i S (p, p') = (2 \pi)^4 \delta^{(4)} (p - p') i S_0 (p)
+  i S_0 (p) {\cal A}_q (p, p') i S_0 (p') \, ,
\end{equation}
which consists of a free term $S_0$ and an interaction part $S_0 {\cal A}_q
S_0$. Making use of the explicit form of the gauge potential
(\ref{Shock-momentum}), the amplitude reads
\begin{eqnarray}
{\cal A}_q (p, p')
\!\!\!&=&\!\!\! 2 \pi \delta (p_- - p'_-) \gamma_-
\sum_{N = 1}^{\infty}
\left( i g \right)^N
\nonumber\\
&&\times\int
\left(
\prod_{j = 1}^{N} \frac{d^2 \mbox{\boldmath$k$}_j}{(2 \pi)^2}
\right)
(2 \pi)^2 \delta^{(2)}
\left(
\mbox{\boldmath$p$}
+
\sum_{l = 1}^{N}
\mbox{\boldmath$k$}_l
-
\mbox{\boldmath$p$}'
\right)
{\cal J}_N (p, p')
\prod_{m = 1}^{N} \beta (\mbox{\boldmath$k$}_m) \, .
\end{eqnarray}
Here
\begin{eqnarray}
{\cal J}_N
= - \frac{i}{N!} \int
\left( \prod_{j = 1}^{N} \frac{d x_j}{2 \pi} \right)
2 \pi
\delta
\left( \sum_{l = 1}^{N} x_l - A \right)
\left( \sum_{m = 1}^{N} x_m \right)
\prod_{n = 1}^{N} \frac{i}{x_n + i 0 \cdot \epsilon (p_-)}
=
\frac{\epsilon^{N - 1}(p_-)}{N !} \, ,
\end{eqnarray}
with $A \equiv p'_+ - \left( \mbox{\boldmath$p$} +
\sum_{j = 1}^{N} \mbox{\boldmath$k$}_j \right)^2/(2 p_-)$ and
the standard sign-function $\epsilon (x) = \theta (x) - \theta (-x)$.

Using this result one immediately finds
\begin{equation}
{\cal A}_q (p, p')
= 2 \pi \delta (p_- - p'_-) \gamma_- \epsilon (p_-)
\int d^2 \mbox{\boldmath$z$} \,
{\rm e}^{- i \mbox{\boldmath${\scriptstyle z}$} \cdot
( \mbox{\boldmath${\scriptstyle p}$}
- \mbox{\boldmath${\scriptstyle p}$}') }
\left\{
u (p_-, \mbox{\boldmath$z$})
- 1
\right\} \, ,
\end{equation}
where we introduced the notation for the color matrix depending on the
external field configuration
\begin{equation}
u {}^i{}_j (p_-, \mbox{\boldmath$z$})
=
{\rm e}^{i g \, \epsilon (p_-)
\beta^i{}_j (\mbox{\boldmath${\scriptstyle z}$})} \, ,
\end{equation}
with the gauge field being a matrix in the fundamental representation
$\beta^i{}_j = \beta^a \left(t^a\right)^i{}_j$. Using the identity
\begin{equation}
\epsilon^{N}(x) = \theta (x) + (- 1)^N \theta (-x) \, ,
\end{equation}
one reduces the propagator to the one already known \cite{Bal96}.

\begin{itemize}
\item Gluon propagator: $- i G^{ab}_{\mu \nu} (x, y) =
\langle 0 | T b^a_\mu (x) b^b_\nu (y) | 0 \rangle_{B}$.
\end{itemize}
The free-field propagator in the light-cone gauge
reads
\begin{equation}
G_{0, \, \mu \nu} (p) = \frac{d_{\mu \nu} (p)}{p^2 + i 0} \, ,
\qquad\qquad\qquad
d_{\mu \nu} (p) = g_{\mu \nu}
- \frac{p_\mu n^\star_\nu + p_\nu n^\star_\mu}{[p_-]} \, ,
\end{equation}
where the square brackets on the $p_-$-pole stand for a particular
prescription to go around it in perturbative computations.
The main advantage of the light-like $b_- = 0$ is the absence of the
quartic $BBbb$ interaction vertices, so that we have only triple
$Bbb$ vertex left. Moreover a simple analysis shows that the only
relevant part of the three-gluon interaction Lagrangian is
\begin{equation}
\delta {\cal L} = - g f^{abc} \left( \partial_\mu b^a_\nu \right)
B^b_\mu b^c_\nu \, ,
\end{equation}
since other contributions, $\sim b_\mu B_\mu$, vanish by virtue of the
orthogonality property of the light-like-gauge gluon propagator
$n^\star_\mu G_{0, \, \mu \nu} = 0$, since $B_\mu \sim n^\star_\mu$.

The manipulations analogous to the one we have done previously with the
fermion Green function give:
\begin{equation}
(- i) G_{\mu \nu} (p, p') = (2 \pi)^4 \delta^{(4)} (p - p')
(- i) G_{0, \, \mu \nu} (p)
+  (- i) G_{0, \, \mu \rho} (p) {\cal A}_{g, \, \rho \sigma} (p, p')
(- i) G_{0, \, \sigma \nu} (p') \, ,
\end{equation}
with
\begin{equation}
\label{GluonA}
{\cal A}_{g, \, \mu \nu} (p, p')
= - 2 \pi \delta (p_- - p'_-) 2 p_- \epsilon (p_-) g_{\mu \nu}
\int d^2 \mbox{\boldmath$z$} \,
{\rm e}^{- i \mbox{\boldmath${\scriptstyle z}$} \cdot
( \mbox{\boldmath${\scriptstyle p}$}
- \mbox{\boldmath${\scriptstyle p}$}') }
\left\{
u (p_-, \mbox{\boldmath$z$})
- 1
\right\} \, ,
\end{equation}
where we used the matrix notation for the gluon field in the adjoint
representation $\beta^{ab} \equiv i f^{acb} \beta^c$ and the adjoint
gauge orientation matrix is related to the ones in the fundamental
representation of the color group via
\begin{equation}
u^{ab} (p_-, \mbox{\boldmath$z$})
=
{\rm e}^{ i g \, \epsilon (p_-)
\beta^{ab} (\mbox{\boldmath${\scriptstyle z}$})}
=
2 {\rm tr}
\left\{
t^a u (p_-, \mbox{\boldmath$z$}) t^b u^\dagger (p_-, \mbox{\boldmath$z$})
\right\} \, .
\end{equation}
Moreover it satisfies the following hermiticity property
\begin{equation}
\label{WLproperty}
u^{\dagger ab} (\mbox{\boldmath$z$}) = u^{ba}
(\mbox{\boldmath$z$}) \, ,
\end{equation}
to be used extensively later. In Eq.\ (\ref{GluonA}) we have also used an
obvious property
\begin{equation}
d_{\mu_1 \mu_2} (p_1) d_{\mu_2 \mu_3} (p_2)
\dots
d_{\mu_{N - 1} \mu_{N}} (p_{N - 1})
=
d_{\mu_1 \mu} (p_1)
d_{\mu \mu_N} (p_{N - 1}) \, .
\end{equation}

Let us briefly discuss several possibilities for handling the spurious
infrared $1/p_-$-pole in the density matrix of the gluon propagator.
Obviously, it is related to the residual gauge degree of freedom: one
can perform a $x_+$-independent gauge transformation which does not
affect the light-cone gauge condition. This ambiguity can be fixed by
imposing a boundary condition in $x_+$ on the gauge field. The vanishing
of the gauge field at $x_+ = \pm \infty$ results into the advanced or
retarded prescription on the pole
\begin{equation}
\frac{1}{[p_-]} = \frac{1}{p_- \pm i 0} \, .
\end{equation}
Their semi-sum results into Cauchy principal value (PV) prescription (which
however does not follow from the path integral quantization). Next, it
can be handled according to the Mandelstam-Leibbrandt (ML) recipe
\cite{Man83,Lei84}
\begin{equation}
\frac{1}{[p_-]} \stackrel{M}{=} \frac{1}{p_- + i 0 \cdot \epsilon (p_+)}
\stackrel{L}{=} \frac{p_+}{p_-p_+ + i 0}
\, ,
\end{equation}
which puts this spurious pole on the same footing as the conventional
pole in the propagator treated by means of causal Feynman prescription.
This recipe allows for the Wick rotation in the Feynman integrals. The ML
form of the propagator does not correspond to simple boundary conditions
on the gauge field at $x_+$-infinity which manifest a residual gauge
freedom. However, this prescription has been deduced later by means of
the canonical equal-time (but not the light-front) quantization
\cite{BasDalLazSol85} and path integral formalism by means of the
Faddeev-Popov trick by changing gauge condition in the path integral
from the temporal to the light-like gauge \cite{SlaFro88}. The ML recipe
results into a mild infrared behaviour as compared to strong singularities
one encounters when the pole is hadled by means of the Cauchy principal
value.

Note, however, that in our present circumstances due to the strip
restriction (\ref{Strip}) on the integration over the $p_-$ component
of teh momentum, the actual pole in the gluon propagator is not hit
and all prescriptions lead to identical results.

\section{Nonlinear evolution equation.}
\label{LOandNLOnonlinearities}

In order to derive the evolution equation for the dipole density, we
develop a perturbation theory for the quantum fields integrating them
out in slices of $p_-$ momentum while keeping the external shock-wave
field to all orders. In the Wilson line formalism, the quark traveling
along the path $x_\mu = x_- n_\mu + x_{\perp\mu}$ interacts with soft
gluons by means of the path-ordered exponential (\ref{PathExp}). For
the gluon field (\ref{ClasQuantDecomp}) it has the form
\begin{eqnarray}
{\cal U} \left( \infty, - \infty; \mbox{\boldmath$x$} \right)
\equiv P \exp \left(
i g \int_{- \infty}^{\infty} dx_- A_+ (x_-, \mbox{\boldmath$x$})
\right)
=
U \left( \infty, 0; \mbox{\boldmath$x$} \right) u (\mbox{\boldmath$x$})
U \left( 0, - \infty; \mbox{\boldmath$x$} \right) \, .
\end{eqnarray}
Here on the right hand side of the equality we have used the form of the
shock-wave concentrated on the plane $x_- = 0$ so that $U$ and $u$ stand
for the path-ordered exponentials containing the quantum and shock-wave
fields, respectively. The quantum Wilson line are to be expanded in
perturbation series, e.g.,
\begin{equation}
U \left( \infty, 0; \mbox{\boldmath$x$} \right)
=
\1 + \sum_{k = 1}^{\infty} (i g)^k \int_{0}^{\infty} d x_{1 -}
\int_{0}^{x_{1 -}} d x_{2 -} \dots \int_{0}^{x_{(k - 1) -}} d x_{k -} \,
b (x_{1 -}, \mbox{\boldmath$x$}) \, b (x_{2 -}, \mbox{\boldmath$x$})
\dots
b (x_{k -}, \mbox{\boldmath$x$}) \, .
\end{equation}
For the hermitian conjugate we have
\begin{equation}
\label{PertExpWilson}
U^\dagger \left( \infty, 0; \mbox{\boldmath$x$} \right)
=
\1 + \sum_{k = 1}^{\infty} (- i g)^k \int_{0}^{\infty} d x_{1 -}
\int_{0}^{x_{1 -}} d x_{2 -} \dots \int_{0}^{x_{(k - 1) -}} d x_{k -} \,
b (x_{k -}, \mbox{\boldmath$x$})
\dots
b (x_{2 -}, \mbox{\boldmath$x$})
\,
b (x_{1 -}, \mbox{\boldmath$x$})
\, .
\end{equation}
so that the unitarity property is preserved
\begin{equation}
\label{Unitarity}
U \left( \infty, 0; \mbox{\boldmath$x$} \right)
U^\dagger \left( \infty, 0; \mbox{\boldmath$x$} \right)
= \1 \, .
\end{equation}
Now we substitute the above expansion into the formula for the dipole
\begin{eqnarray*}
{\cal N} (\mbox{\boldmath$x$}, \mbox{\boldmath$y$})
= \langle 0 | T \, \frac{1}{N_c} {\rm tr}
\left(
{\cal U} \left( \infty, - \infty; \mbox{\boldmath$x$} \right)
{\cal U}^\dagger \left( \infty, - \infty; \mbox{\boldmath$y$} \right)
- \1
\right)
| 0 \rangle \, ,
\end{eqnarray*}
and form Wick contractions. Note, that without adhering to the large $N_c$
approximation our evolution equation will not be closed but rather it will
involve the higher correlations of Wilson lines, e.g.,
\begin{eqnarray}
{\cal N}_{(2)}
(\mbox{\boldmath$x$}, \mbox{\boldmath$z$}, \mbox{\boldmath$y$})
\!\!\!&=&\!\!\! \langle 0 | T \,
\frac{1}{N_c} {\rm tr}
\left(
{\cal U} \left( \infty, - \infty; \mbox{\boldmath$x$} \right)
{\cal U}^\dagger \left( \infty, - \infty; \mbox{\boldmath$z$} \right)
- \1
\right) \nonumber\\
&&\quad\times
\frac{1}{N_c} {\rm tr}
\left(
{\cal U} \left( \infty, - \infty; \mbox{\boldmath$z$} \right)
{\cal U}^\dagger \left( \infty, - \infty; \mbox{\boldmath$y$} \right)
- \1
\right)
| 0 \rangle \, ,
\end{eqnarray}
etc. However, the multicolor limit will give a possibility to reduce
all higher order correlation to a simple product of dipole densities,
see e.g., Eq.\ (\ref{N2toNNlargeNc}) below.

\section{Leading nonlinearities.}

\begin{figure}[t]
\begin{center}
\hspace{0cm} \mbox{
\begin{picture}(0,93)(100,0)
\put(10,0){\insertfig{6}{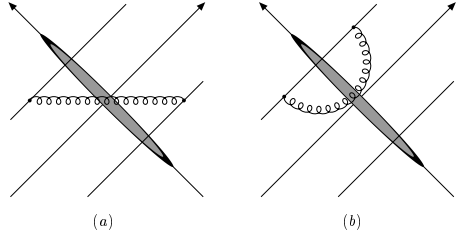}}
\end{picture}
}
\end{center}
\caption{\label{lo} Leading order diagrams which result into BFKL
equation and the first nonlinear correction to the generalized
nonlinear equation.}
\end{figure}

To start with let us recapitulate the computation of the BFKL part and
leading nonlinearities of the generalized equation (\ref{GeneralNonlinear}).
To this end let us compute the diagrams given in Fig.\ \ref{lo}. For these
contributions we only need the $++$-projection of the shock-wave propagator
which reads
\begin{eqnarray}
G_{++} (p, p')
\!\!\!&=&\!\!\! - (2 \pi)^4 \delta^{(4)} (p - p')
\frac{2}{p^2 + i 0} \frac{p_+}{[p_-]}
\nonumber\\
&-&\!\!\! 2 \pi \, \delta (p_- - p'_-) \frac{2 i}{(p^2 + i 0) (p'^{2} + i 0)}
\frac{\mbox{\boldmath$p$} \cdot \mbox{\boldmath$p$}'}{[p_-]}
\int d^2 \mbox{\boldmath$z$} \,
{\rm e}^{- i \mbox{\boldmath${\scriptstyle z}$} \cdot
( \mbox{\boldmath${\scriptstyle p}$}
- \mbox{\boldmath${\scriptstyle p}$}') }
\nonumber\\
&&\times
\left\{
\theta (p_-) \left( u (\mbox{\boldmath$z$}) - 1 \right)
-
\theta (- p_-) \left( u^\dagger (\mbox{\boldmath$z$}) - 1 \right)
\right\} \, .
\end{eqnarray}

We can cast the free propagator to the form when the particle propagate
to an intermediate point $\mbox{\boldmath$z$}$ before it reaches its
final destination, a form we have for the interaction part with the shock-wave
background. To achieve this, we integrate, in the Fourier transform of the
free light-cone-gauge propagator, over the $p_+$-component, then insert the
unity $1 = \int d^2 \mbox{\boldmath$p$}' \delta^{(2)} (\mbox{\boldmath$p$}
- \mbox{\boldmath$p$}')$, and use the integral representation of the
$\delta$-function. These manipulations lead to the result
\begin{eqnarray}
\label{Gluon++}
G_{0, ++} (x - y)
\!\!\!&=&\!\!\! - \int \frac{d^4 p}{(2 \pi)^4} \, {\rm e}^{- i p \cdot (x - y)}
\frac{2}{p^2 + i 0} \frac{p_+}{[p_-]} \\
&=&\!\!\!
\frac{i}{2} \int \frac{d p_-}{2 \pi} \frac{1}{[p_-]}
\left\{
\theta (x_- - y_-) \theta (p_-)
-
\theta (y_- - x_-) \theta (-p_-)
\right\}
{\rm e}^{- i p_- (x - y)_+}
\nonumber\\
&&\times \int d^2 \mbox{\boldmath$z$}
\int \frac{d^2 \mbox{\boldmath$p$}}{(2 \pi)^2}
{\rm e}^{i \mbox{\boldmath${\scriptstyle p}$} \cdot
( \mbox{\boldmath${\scriptstyle x}$} - \mbox{\boldmath${\scriptstyle z}$} )
- i \mbox{\boldmath${\scriptstyle p}$}^2/(2 p_-) x_-}
\int \frac{d^2 \mbox{\boldmath$p$}'}{(2 \pi)^2}
{\rm e}^{- i \mbox{\boldmath${\scriptstyle p}$}' \cdot
( \mbox{\boldmath${\scriptstyle y}$} - \mbox{\boldmath${\scriptstyle z}$} )
+ i \mbox{\boldmath${\scriptstyle p}$}'^2/(2 p_-) y_-} \,
\frac{\mbox{\boldmath$p$} \cdot \mbox{\boldmath$p$}'}{(p_-)^2}
\nonumber\\
&&-
\delta (x_- - y_-)
\delta^{(2)}
( \mbox{\boldmath$x$} - \mbox{\boldmath$y$} )
\int \frac{d p_-}{2 \pi} \frac{1}{[p_-] p_-}
{\rm e}^{- i p_- (x - y)_+}
\, ,
\end{eqnarray}
where we have used a $p_+$-independent regularization of the spurious
$1/[p_-]$-pole, e.g., the advanced/retarded/PV prescription, not the ML
which does depend on $p_+$. However, since $p_-$ integration is resctricted
this limitation is irrelevant. In Eq.\ (\ref{Gluon++}) the last term
originates from the `Coulomb force' \cite{KogSop70}.

Note that if the integration over $p_-$ would be unrestricted then
one will use a property of the propagator with ML prescription
which follows from its causality and conclude that
\begin{equation}
G_{0, ++} ( x_+ = 0, \mbox{\boldmath$x$}, x_- ) \stackrel{ML}{=} 0 \, ,
\end{equation}
which holds when one performs the $p_-$ integral in the complex plane
and notices that due to the ML prescription both, Feynman and spurious,
poles lie on the same side of the real $p_-$ axis. This would lead
in turn to the inability to reproduce the BFKL equation. This will hold
even if one would use the regularization by means of displacing the path
from the light-like path adding a deviation $\delta x_+$ since the
corresponding contributions are finite, they do not contain double
logarithmic divergence as in the principal value prescription.

Note that since we have the integration over $p_-$ in the restricted
domain (\ref{Strip}),
\begin{eqnarray*}
\int_{p_{0-}}^{p_{0-} + \delta p_-} \frac{d p_-}{p_-}
= \ln \left( \frac{p_{0-} + \delta p_-}{p_{0-}} \right)
= \ln\frac{1}{x_{\rm B}} \, ,
\end{eqnarray*}
we never hit the `spurious' pole in the gluon propagator in the one-loop
diagrams since the `ghost' part of the propagator is concentrated at $p_- = 0$,
namely,
\begin{eqnarray*}
\frac{p_+}{p_- p_+ + i 0} = {\rm PV} \frac{1}{p_-} - i \pi
\epsilon(p_+) \delta (p_-) \, .
\end{eqnarray*}
Therefore, to the leading order in the coupling constant all prescriptions
on the $p_-$-pole are equivalent, as we already mentioned above.

Adding the free and interaction pieces together we get
\begin{eqnarray}
G_{++} (x, y)
\!\!\!&=&\!\!\!
\frac{i}{2} \int \frac{d p_-}{2 \pi} \frac{1}{[p_-]}
\left\{
\theta (x_- - y_-) \theta (p_-)
-
\theta (y_- - x_-) \theta (-p_-)
\right\}
{\rm e}^{- i p_- (x - y)_+} \\
&&\times \int d^2 \mbox{\boldmath$z$}
\int \frac{d^2 \mbox{\boldmath$p$}}{(2 \pi)^2}
{\rm e}^{i \mbox{\boldmath${\scriptstyle p}$} \cdot
( \mbox{\boldmath${\scriptstyle x}$} - \mbox{\boldmath${\scriptstyle z}$} )
- i \mbox{\boldmath${\scriptstyle p}$}^2/(2 p_-) x_-}
\int \frac{d^2 \mbox{\boldmath$p$}'}{(2 \pi)^2}
{\rm e}^{- i \mbox{\boldmath${\scriptstyle p}$}' \cdot
( \mbox{\boldmath${\scriptstyle y}$} - \mbox{\boldmath${\scriptstyle z}$} )
+ i \mbox{\boldmath${\scriptstyle p}$}'^2/(2 p_-) y_-} \,
\frac{\mbox{\boldmath$p$} \cdot \mbox{\boldmath$p$}'}{(p_-)^2}
\nonumber\\
&&\times
\left\{
\theta (x_-) \theta (y_-) + \theta (- x_-) \theta (- y_-)
+ \theta (x_-) \theta (- y_-) u (\mbox{\boldmath$z$})
+ \theta (- x_-) \theta (y_-) u^\dagger (\mbox{\boldmath$z$})
\right\}
\nonumber\\
&&-
\delta (x_- - y_-)
\int \frac{d p_-}{2 \pi} \frac{1}{[p_-]}
{\rm e}^{- i p_- (x - y)_+}
\int \frac{d^2 \mbox{\boldmath$p$}}{(2 \pi)^2}
{\rm e}^{i \mbox{\boldmath${\scriptstyle p}$} \cdot
( \mbox{\boldmath${\scriptstyle x}$} - \mbox{\boldmath${\scriptstyle y}$} )}
\,
\frac{\mbox{\boldmath$p$}^2}{p_-} \, .
\end{eqnarray}

Since the computation of the one-loop graphs is trivial we just mention
that we have used in the derivation the Euclidean $d$-dimensional transverse
space Fourier transformation,
\begin{equation}
\label{EuclidInt}
\int \frac{d^d \mbox{\boldmath$p$}}{(2 \pi)^d}
\frac{
{\rm e}^{i \mbox{\boldmath${\scriptstyle p}$} \cdot
\mbox{\boldmath${\scriptstyle z}$} }
}{
\mbox{\boldmath$p$}^{2m}}
= \frac{1}{2^{2 m} \pi^{d/2}}
\frac{{\mit\Gamma} (d/2 - m)}{{\mit\Gamma} (m)}
\frac{1}{\mbox{\boldmath$z$}^{d - 2m}} \, ,
\end{equation}
and the following identity
\begin{equation}
2 u^{ab} (\mbox{\boldmath$z$})
{\rm tr}
\left\{
t^a u (\mbox{\boldmath$x$}) t^b u^\dagger (\mbox{\boldmath$y$})
\right\}
=
{\rm tr}
\left\{
u (\mbox{\boldmath$z$}) u^\dagger (\mbox{\boldmath$y$})
\right\}
{\rm tr}
\left\{
u (\mbox{\boldmath$x$}) u^\dagger (\mbox{\boldmath$z$})
\right\}
-
\frac{1}{N_c}
{\rm tr}
\left\{
u (\mbox{\boldmath$x$}) u^\dagger (\mbox{\boldmath$y$})
\right\} \, ,
\end{equation}
stemming from the color Fiertz transformation.

The contribution of the exchange-type diagrams (with one of them sampled
in Fig.\ \ref{lo} ($a$)) reads
\begin{eqnarray}
\label{LOdiagA}
{\cal N} (\mbox{\boldmath$x$}, \mbox{\boldmath$y$})
\!\!\!&=&\!\!\! \frac{\alpha_s}{2\pi^2} N_c \ln\frac{1}{x_{\rm B}}
\int d^2 \mbox{\boldmath$z$}
(-2)
\frac{( \mbox{\boldmath$z$} - \mbox{\boldmath$x$} )
\cdot ( \mbox{\boldmath$z$} - \mbox{\boldmath$y$} )}
{( \mbox{\boldmath$z$} - \mbox{\boldmath$x$} )^2
( \mbox{\boldmath$z$} - \mbox{\boldmath$y$} )^2}
 \nonumber\\
&&\qquad\qquad\qquad\quad\times
\left\{
{\cal N} (\mbox{\boldmath$x$}, \mbox{\boldmath$z$})
+
{\cal N} (\mbox{\boldmath$z$}, \mbox{\boldmath$y$})
-
{\cal N} (\mbox{\boldmath$x$}, \mbox{\boldmath$y$})
+
{\cal N}_{(2)}
(\mbox{\boldmath$x$}, \mbox{\boldmath$z$}, \mbox{\boldmath$y$})
\right\} \, .
\end{eqnarray}
The contribution of the diagram \ref{lo} ($b$) (and analogous one with the
self-energy attached to the other eikonal line) is
\begin{eqnarray}
\label{LOdiagB}
{\cal N} (\mbox{\boldmath$x$}, \mbox{\boldmath$y$})
\!\!\!&=&\!\!\! \frac{\alpha_s}{2\pi^2} N_c \ln\frac{1}{x_{\rm B}}
\int d^2 \mbox{\boldmath$z$}
\left\{
\frac{1}{
( \mbox{\boldmath$z$} - \mbox{\boldmath$x$} )^2}
+
\frac{1}{
( \mbox{\boldmath$z$} - \mbox{\boldmath$y$} )^2}
\right\} \nonumber\\
&&\qquad\qquad\qquad\quad\times
\left\{
{\cal N} (\mbox{\boldmath$x$}, \mbox{\boldmath$z$})
+
{\cal N} (\mbox{\boldmath$z$}, \mbox{\boldmath$y$})
-
{\cal N} (\mbox{\boldmath$x$}, \mbox{\boldmath$y$})
+
{\cal N}_{(2)}
(\mbox{\boldmath$x$}, \mbox{\boldmath$z$}, \mbox{\boldmath$y$})
\right\} \, .
\end{eqnarray}
The `Coulomb' piece does not produce the logarithmic integral at leading
order and, moreover, can be completely eliminated at this order by a
regularization of the Wilson line enforcing the ordering of emitted gluons
by a infinitezimal cutoff, so that e.g., $x_2 \leq x_1 - 0_+$ in Eq.\
(\ref{PertExpWilson}).

In these computations no multicolor approximation has been involved.
However, to produce the closed equation for dipole densities we have to adhere
to the large $N_c$ limit,
\begin{eqnarray}
\label{N2toNNlargeNc}
{\cal N}_{(2)}
(\mbox{\boldmath$x$}, \mbox{\boldmath$z$}, \mbox{\boldmath$y$})
=
{\cal N} (\mbox{\boldmath$x$}, \mbox{\boldmath$z$})
{\cal N} (\mbox{\boldmath$z$}, \mbox{\boldmath$y$}) \, .
\end{eqnarray}

Summing up the expressions (\ref{LOdiagA}) and (\ref{LOdiagB}) together with
the approximation (\ref{N2toNNlargeNc}) we reduce the result to the nonlinear
equation
\begin{eqnarray}
\frac{d}{d \ln \frac{1}{x_{\rm B}}}
{\cal N} (\mbox{\boldmath$x$}, \mbox{\boldmath$y$})
\!\!\!&=&\!\!\! \frac{\alpha_s}{2\pi^2} N_c
\int d^2 \mbox{\boldmath$z$}
\frac{( \mbox{\boldmath$x$} - \mbox{\boldmath$y$} )^2 }
{( \mbox{\boldmath$z$} - \mbox{\boldmath$x$} )^2
( \mbox{\boldmath$z$} - \mbox{\boldmath$y$} )^2} \nonumber\\
&&\qquad\qquad\quad\times
\left\{
{\cal N} (\mbox{\boldmath$x$}, \mbox{\boldmath$z$})
+
{\cal N} (\mbox{\boldmath$z$}, \mbox{\boldmath$y$})
-
{\cal N} (\mbox{\boldmath$x$}, \mbox{\boldmath$y$})
+
{\cal N}
(\mbox{\boldmath$x$}, \mbox{\boldmath$z$})
{\cal N}
(\mbox{\boldmath$z$}, \mbox{\boldmath$y$})
\right\} \, ,
\end{eqnarray}
which is obviously Eq.\ (\ref{GeneralNonlinear}) with ${\cal K}_{1,2}$ given
by Eqs.\ (\ref{BFKLkernel}) and (\ref{LOnonLkernel}), respectively.

\section{Next-to-leading nonlinearities.}

Now we are in a position to address the next-to-leading nonlinearities
to the generalized nonlinear equation. To this end we have to evaluate
the graphs shown in Fig.\ \ref{nlo}. In their computation we will keep
only the contributions which are bilinear in $u^{ab}$ stemming from
the gluon propagators in the external field and omit the color
suppressed terms after application of Fiertz identities. Thus we will
not endeavor the calculation of the radiative corrections to the
leading results which arise from the same diagrams but with only one
or none internal gluons intersecting the shock-wave.

\subsection{Preliminaries.}

Beyond leading order all components off the gluon propagator are relevant.
Therefore, we find it instructive to use the Sudakov decomposition for
momentum and decompose the density matrix into the part orthogonal to
the vector $n^\star$ and the rest,
\begin{equation}
d_{\mu \nu} (p) \equiv
d^\perp_{\mu \nu} (p)
- 2 \frac{p_+}{[p_-]} n^\star_\mu n^\star_\nu
= g^\perp_{\mu \nu}
- \frac{p^\perp_\mu n^\star_\nu + p^\perp_\nu n^\star_\mu}{[p_-]}
- 2 \frac{p_+}{[p_-]} n^\star_\mu n^\star_\nu \, .
\end{equation}
Obviously, $d_{\mu \nu} (p, p_+ = 0) \equiv d^\perp_{\mu \nu} (p)$.
Note that the last term is relevant for the free propagator only since
it vanishes in the interacting piece as the $d$'s are contracted in an
index $d_{\mu \rho} (p) d_{\rho \nu} (p') = d^\perp_{\mu \rho} (p)
d^\perp_{\rho \nu} (p')$.

Similarly to the treatment of $++$ component of the gauge propagator
in the preceding section, we find that the most convenient form of the
latter is to integrate out the `$+$'-momenta in the $d^\perp$-part.
This results into the on-shell condition for the virtuality,
$p_+ = \mbox{\boldmath$p$}^2/(2p_-)$. In this way, we have for the free
propagator,
\begin{eqnarray}
\label{FreePropagator}
G_{0, \mu \nu} (x - y) \!\!\!&=&\!\!\!
- \frac{i}{2} \int \frac{d p_-}{2 \pi}
\left\{
\theta (x_- - y_-) \theta (p_-)
-
\theta (y_- - x_-) \theta (- p_-)
\right\}
{\rm e}^{- i p_- (x - y)_+}
\\
&\times&\!\!\!
\int \frac{d^2 \mbox{\boldmath$p$}}{(2 \pi)^2}
{\rm e}^{i \mbox{\boldmath${\scriptstyle p}$} \cdot
( \mbox{\boldmath${\scriptstyle x}$} - \mbox{\boldmath${\scriptstyle y}$} )
- i \mbox{\boldmath${\scriptstyle p}$}^2/(2 p_-) ( x_- - y_- )}
\frac{1}{p_-} d^\perp_{\mu\nu} (p)
- 2 n^\star_\mu n^\star_\nu
\int \frac{d^4 p}{(2 \pi)^4} {\rm e}^{- i p \cdot (x - y)}
\frac{1}{p^2 + i 0} \frac{p_+}{[p_-]}
\, . \nonumber
\end{eqnarray}
and for the part interacting with the shock-wave background
\begin{eqnarray}
\label{InteractionPropagator}
G^I_{\mu \nu} (x, y) \!\!\!&=&\!\!\!
- \frac{i}{2} \int \frac{d p_-}{2 \pi} {\rm e}^{- i p_- (x - y)_+}
\int d^2 \mbox{\boldmath$z$}
\left\{
\theta (x_-) \theta(- y_-) \theta (p_-)
u (\mbox{\boldmath$z$})
-
\theta (- x_-) \theta(y_-) \theta (- p_-)
u^\dagger (\mbox{\boldmath$z$})
\right\}
\nonumber\\
&&\times
\int \frac{d^2 \mbox{\boldmath$p$}}{(2 \pi)^2}
{\rm e}^{i \mbox{\boldmath${\scriptstyle p}$} \cdot
( \mbox{\boldmath${\scriptstyle x}$} - \mbox{\boldmath${\scriptstyle z}$} )
- i \mbox{\boldmath${\scriptstyle p}$}^2/(2 p_-) x_-}
\int \frac{d^2 \mbox{\boldmath$p$}'}{(2 \pi)^2}
{\rm e}^{ - i \mbox{\boldmath${\scriptstyle p}$}' \cdot
( \mbox{\boldmath${\scriptstyle y}$} - \mbox{\boldmath${\scriptstyle z}$} )
+ i \mbox{\boldmath${\scriptstyle p}$}'^2/(2 p_-) y_- }
\frac{1}{p_-} d^\perp_{\mu\rho} (p) d^\perp_{\rho\nu} (p') \, .
\end{eqnarray}

\begin{figure}[t]
\begin{center}
\hspace{0cm} \mbox{
\begin{picture}(0,313)(100,0)
\put(-100,0){\insertfig{14}{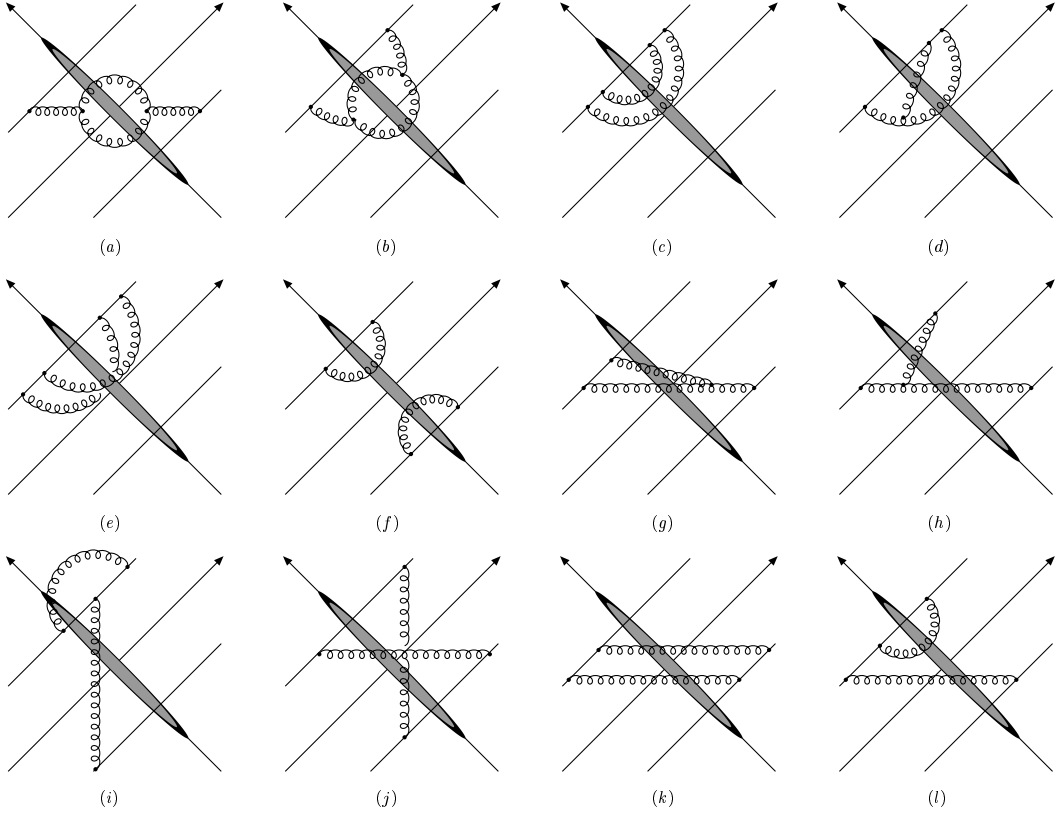}}
\end{picture}
}
\end{center}
\caption{\label{nlo} Samples of diagrams of different topologies
inducing the next-to-leading nonlinearity, cubic in dipole densities,
in the generalized nonlinear equation.}
\end{figure}

Using these rules together with the conventional expressions for
the QCD interaction vertices we are in a position to compute the
graphs displayed in Fig.\ \ref{nlo}. We divide the diagrams into
graphs of different topologies, which we name self-energy ($a$-$f$),
vertex ($g$-$i$) and box ($j$-$l$) types. Although this division is
ambiguous we stick to it for the rest of this section.

\subsection{Self-energy topology.}

The calculation of the gluon bubble in the external field leads to the
most involved algebra. Leaving technical details of this exercise to
Appendix \ref{NonAbelianDiag}, we give here just the final result for
the diagram \ref{nlo} ($a$) which reads
\begin{eqnarray}
\label{DiagA}
{\cal N} (\mbox{\boldmath$x$}, \mbox{\boldmath$y$})
\!\!\!&=&\!\!\!
- 2 \frac{\alpha_s^2}{N_c} \int_{0}^{1} du
\int \frac{d p_-}{p_-}
\int d^2 \mbox{\boldmath$z$} \ d^2 \mbox{\boldmath$z$}'
\
{\rm tr}
\left\{
\left(
t^a
u (\mbox{\boldmath$x$})
t^d
+
t^d
u (\mbox{\boldmath$x$})
t^a
\right)
u^\dagger (\mbox{\boldmath$y$})
\right\}
\, f^{abc} f^{def} \,
u^{be} (\mbox{\boldmath$z$})
u^{cf} (\mbox{\boldmath$z$}')
\nonumber\\
&\times&\!\!\!
\int \frac{d^2 \mbox{\boldmath$p$}_1}{(2 \pi)^2}
\frac{d^2 \mbox{\boldmath$p$}'_1}{(2 \pi)^2}
\frac{d^2 \mbox{\boldmath$p$}_2}{(2 \pi)^2}
\frac{d^2 \mbox{\boldmath$p$}'_2}{(2 \pi)^2}
\frac{
{\rm e}^{i ( \mbox{\boldmath${\scriptstyle p}$}_1
+ \mbox{\boldmath${\scriptstyle p}$}_2) \cdot
\mbox{\boldmath${\scriptstyle x}$}
- i ( \mbox{\boldmath${\scriptstyle p}$}'_1
+ \mbox{\boldmath${\scriptstyle p}$}'_2) \cdot
\mbox{\boldmath${\scriptstyle y}$}
- i ( \mbox{\boldmath${\scriptstyle p}$}_1
- \mbox{\boldmath${\scriptstyle p}$}'_1 ) \cdot
\mbox{\boldmath${\scriptstyle z}$}
- i ( \mbox{\boldmath${\scriptstyle p}$}_2
- \mbox{\boldmath${\scriptstyle p}$}'_2 ) \cdot
\mbox{\boldmath${\scriptstyle z}$}'}
}{
\left( \mbox{\boldmath$p$}_1 + \mbox{\boldmath$p$}_2 \right)^2
\left( \mbox{\boldmath$p$}'_1 + \mbox{\boldmath$p$}'_2 \right)^2
\left( \bar u \mbox{\boldmath$p$}_1^2 + u \mbox{\boldmath$p$}_2^2 \right)
\left( \bar u {\mbox{\boldmath$p$}'_1}^2 + u {\mbox{\boldmath$p$}'_2}^2 \right)
}
\nonumber\\
&&\qquad\times
\Bigg\{ u (1 + u)
\Big\{
\left(
\mbox{\boldmath$p$}_1^2 - \mbox{\boldmath$p$}_2^2
\right)
\
\mbox{\boldmath$p$}'_2
\cdot
\mbox{\boldmath$p$}'
+
\left(
{\mbox{\boldmath$p$}'_1}^2 - {\mbox{\boldmath$p$}'_2}^2
\right)
\
\mbox{\boldmath$p$}_2
\cdot
\mbox{\boldmath$p$}
+
2 \,
\mbox{\boldmath$p$}_2
\cdot
\mbox{\boldmath$p$}'
\
\mbox{\boldmath$p$}'_2
\cdot
\mbox{\boldmath$p$}
\Big\}
\nonumber\\
&&\qquad \ \
+ \bar u (1 + \bar u)
\Big\{
\left(
\mbox{\boldmath$p$}_2^2 - \mbox{\boldmath$p$}_1^2
\right)
\
\mbox{\boldmath$p$}'_1
\cdot
\mbox{\boldmath$p$}'
+
\left(
{\mbox{\boldmath$p$}'_2}^2 - {\mbox{\boldmath$p$}'_1}^2
\right)
\
\mbox{\boldmath$p$}_1
\cdot
\mbox{\boldmath$p$}
+
2 \,
\mbox{\boldmath$p$}_1
\cdot
\mbox{\boldmath$p$}'
\
\mbox{\boldmath$p$}'_1
\cdot
\mbox{\boldmath$p$}
\Big\}
\nonumber\\
&-&\!\!\! 2 (1 + u \bar u)
\Big\{
\mbox{\boldmath$p$}_2
\cdot
\mbox{\boldmath$p$}'
\
\mbox{\boldmath$p$}'_1
\cdot
\mbox{\boldmath$p$}
+
\mbox{\boldmath$p$}_1
\cdot
\mbox{\boldmath$p$}'
\
\mbox{\boldmath$p$}'_2
\cdot
\mbox{\boldmath$p$}
\Big\}
- 2 (1 - u \bar u)
\
\mbox{\boldmath$p$}
\cdot
\mbox{\boldmath$p$}'
\left(
\mbox{\boldmath$p$}_1
\cdot
\mbox{\boldmath$p$}'_2
+
\mbox{\boldmath$p$}_2
\cdot
\mbox{\boldmath$p$}'_1
\right)
\nonumber\\
&&\qquad \ \ + \, \frac{2}{u \bar u}
\
\mbox{\boldmath$p$}
\cdot
\mbox{\boldmath$p$}'
\
\Big\{
u^2 (1 + u^2)
\mbox{\boldmath$p$}_2 \cdot \mbox{\boldmath$p$}'_2
+
\bar u^2 (1 + \bar u^2)
\mbox{\boldmath$p$}_1 \cdot \mbox{\boldmath$p$}'_1
\Big\}
\Bigg\}
\, .
\end{eqnarray}
Here
\begin{equation}
\mbox{\boldmath$p$}
=
\mbox{\boldmath$p$}_1 + \mbox{\boldmath$p$}_2 \, , \qquad
\mbox{\boldmath$p$}'
=
\mbox{\boldmath$p$}'_1 + \mbox{\boldmath$p$}'_2 \, .
\end{equation}
The diagram in Fig.\ \ref{nlo} ($b$) is easily obtained from ($a$) by
keeping only the first term in the color trace, changing the sign
of the whole contribution, and by identifying $\mbox{\boldmath$y$} \to
\mbox{\boldmath$x$}$ in the exponential in the integrand, i.e.,
\begin{equation}
- {\rm e}^{i ( \mbox{\boldmath${\scriptstyle p}$}_1
+ \mbox{\boldmath${\scriptstyle p}$}_2) \cdot
\mbox{\boldmath${\scriptstyle x}$}
- i ( \mbox{\boldmath${\scriptstyle p}$}'_1
+ \mbox{\boldmath${\scriptstyle p}$}'_2) \cdot
\mbox{\boldmath${\scriptstyle x}$}
- i ( \mbox{\boldmath${\scriptstyle p}$}_1
- \mbox{\boldmath${\scriptstyle p}$}'_1 ) \cdot
\mbox{\boldmath${\scriptstyle z}$}
- i ( \mbox{\boldmath${\scriptstyle p}$}_2
- \mbox{\boldmath${\scriptstyle p}$}'_2 ) \cdot
\mbox{\boldmath${\scriptstyle z}$}'} \, .
\end{equation}
When the gluon self-energy is attached to another Wilson line we keep the
second contribution in the color trace, set $\mbox{\boldmath$x$} \to
\mbox{\boldmath$y$}$, and multiply the result by the minus sign, i.e.,
\begin{equation}
- {\rm e}^{i ( \mbox{\boldmath${\scriptstyle p}$}_1
+ \mbox{\boldmath${\scriptstyle p}$}_2) \cdot
\mbox{\boldmath${\scriptstyle y}$}
- i ( \mbox{\boldmath${\scriptstyle p}$}'_1
+ \mbox{\boldmath${\scriptstyle p}$}'_2) \cdot
\mbox{\boldmath${\scriptstyle y}$}
- i ( \mbox{\boldmath${\scriptstyle p}$}_1
- \mbox{\boldmath${\scriptstyle p}$}'_1 ) \cdot
\mbox{\boldmath${\scriptstyle z}$}
- i ( \mbox{\boldmath${\scriptstyle p}$}_2
- \mbox{\boldmath${\scriptstyle p}$}'_2 ) \cdot
\mbox{\boldmath${\scriptstyle z}$}'} \, .
\end{equation}

The two-loop quark self-energy diagrams \ref{nlo} ($c$), ($d$), ($e$)
and ($f$) are computed along the same line. The diagram ($c$) gives
\begin{eqnarray}
{\cal N} (\mbox{\boldmath$x$}, \mbox{\boldmath$y$})
\!\!\!&=&\!\!\!
16 \frac{\alpha_s^2}{N_c} \int_{0}^{1} du
\int \frac{d p_-}{p_-}
\int d^2 \mbox{\boldmath$z$} \ d^2 \mbox{\boldmath$z$}'
\
{\rm tr}
\left\{
t^a t^b
u (\mbox{\boldmath$x$})
t^c t^d
u^\dagger (\mbox{\boldmath$y$})
\right\}
\,
u^{ad} (\mbox{\boldmath$z$})
u^{bc} (\mbox{\boldmath$z$}')
\\
&&\times
\int \frac{d^2 \mbox{\boldmath$p$}_1}{(2 \pi)^2}
\frac{d^2 \mbox{\boldmath$p$}'_1}{(2 \pi)^2}
\frac{d^2 \mbox{\boldmath$p$}_2}{(2 \pi)^2}
\frac{d^2 \mbox{\boldmath$p$}'_2}{(2 \pi)^2}
\frac{\mbox{\boldmath$p$}_1 \cdot \mbox{\boldmath$p$}'_1
\
\mbox{\boldmath$p$}_2
\cdot
\mbox{\boldmath$p$}'_2
}{
\left( \bar u \mbox{\boldmath$p$}_1^2 + u \mbox{\boldmath$p$}_2^2 \right)
\left( \bar u {\mbox{\boldmath$p$}'_1}^2 + u {\mbox{\boldmath$p$}'_2}^2 \right)
} \nonumber\\
&\times&\!\!\!
\Bigg\{
{\rm e}^{i ( \mbox{\boldmath${\scriptstyle p}$}_1
- \mbox{\boldmath${\scriptstyle p}$}'_1 )\cdot
( \mbox{\boldmath${\scriptstyle x}$}
- \mbox{\boldmath${\scriptstyle z}$} )
+
i ( \mbox{\boldmath${\scriptstyle p}$}_2
- \mbox{\boldmath${\scriptstyle p}$}'_2 )\cdot
( \mbox{\boldmath${\scriptstyle x}$}
- \mbox{\boldmath${\scriptstyle z}$}' )
}
\frac{u}{\bar u} \,
\frac{1}{ \mbox{\boldmath$p$}_1^2 \ {\mbox{\boldmath$p$}'_1}^2 }
+ \
{\rm e}^{i ( \mbox{\boldmath${\scriptstyle p}$}_1
- \mbox{\boldmath${\scriptstyle p}$}'_1 )\cdot
( \mbox{\boldmath${\scriptstyle y}$}
- \mbox{\boldmath${\scriptstyle z}$} )
+
i ( \mbox{\boldmath${\scriptstyle p}$}_2
- \mbox{\boldmath${\scriptstyle p}$}'_2 )\cdot
( \mbox{\boldmath${\scriptstyle y}$}
- \mbox{\boldmath${\scriptstyle z}$}' )
}
\frac{\bar u}{u} \,
\frac{1}{ \mbox{\boldmath$p$}_2^2 \ {\mbox{\boldmath$p$}'_2}^2 }
\Bigg\}
\, . \nonumber
\end{eqnarray}
The diagram ($d$) can be deduced from ($g$), to be computed below,
and reads
\begin{eqnarray}
{\cal N} (\mbox{\boldmath$x$}, \mbox{\boldmath$y$})
\!\!\!&=&\!\!\!
8 \frac{\alpha_s^2}{N_c} \int_{0}^{1} du
\int \frac{d p_-}{p_-}
\int d^2 \mbox{\boldmath$z$} \ d^2 \mbox{\boldmath$z$}'
\,
u^{ad} (\mbox{\boldmath$z$})
u^{be} (\mbox{\boldmath$z$}')
\\
&&\times \int \frac{d^2 \mbox{\boldmath$p$}_1}{(2 \pi)^2}
\frac{d^2 \mbox{\boldmath$p$}'_1}{(2 \pi)^2}
\frac{d^2 \mbox{\boldmath$p$}_2}{(2 \pi)^2}
\frac{d^2 \mbox{\boldmath$p$}'_2}{(2 \pi)^2}
\frac{
{\rm e}^{- i ( \mbox{\boldmath${\scriptstyle p}$}_1
- \mbox{\boldmath${\scriptstyle p}$}'_1 ) \cdot
\mbox{\boldmath${\scriptstyle z}$}
- i ( \mbox{\boldmath${\scriptstyle p}$}_2
- \mbox{\boldmath${\scriptstyle p}$}'_2 ) \cdot
\mbox{\boldmath${\scriptstyle z}$}'}
}
{
\left( \mbox{\boldmath$p$}'_1 + \mbox{\boldmath$p$}'_2 \right)^2
\left( \bar u \mbox{\boldmath$p$}_1^2 + u \mbox{\boldmath$p$}_2^2 \right)
\left( \bar u {\mbox{\boldmath$p$}'_1}^2 + u {\mbox{\boldmath$p$}'_2}^2 \right)
}
\nonumber\\
&&\times
\left\{
u \bar u \, \mbox{\boldmath$p$}_1 \cdot \mbox{\boldmath$p$}_2 \
\left( {\mbox{\boldmath$p$}'_1}^2 - {\mbox{\boldmath$p$}'_2}^2 \right)
- 2 \bar u \, \mbox{\boldmath$p$}_1 \cdot \mbox{\boldmath$p$}'_1
\ \mbox{\boldmath$p$}_2 \cdot \left( \mbox{\boldmath$p$}'_1
+ \mbox{\boldmath$p$}'_2 \right)
+ 2 u \, \mbox{\boldmath$p$}_2 \cdot \mbox{\boldmath$p$}'_2
\ \mbox{\boldmath$p$}_1 \cdot \left( \mbox{\boldmath$p$}'_1
+ \mbox{\boldmath$p$}'_2 \right)
\right\}
\nonumber\\
&\times&\!\!\!
\Bigg\{
i f^{cde} \,
{\rm tr}
\left\{
t^a t^b
u (\mbox{\boldmath$x$})
t^c
u^\dagger (\mbox{\boldmath$y$})
\right\}
\left(
\frac{1}{\bar u \mbox{\boldmath$p$}_1^2}
{\rm e}^{i ( \mbox{\boldmath${\scriptstyle p}$}_1
+ \mbox{\boldmath${\scriptstyle p}$}_2
- \mbox{\boldmath${\scriptstyle p}$}'_1
- \mbox{\boldmath${\scriptstyle p}$}'_2 ) \cdot
\mbox{\boldmath${\scriptstyle x}$}}
-
\frac{1}{u \mbox{\boldmath$p$}_2^2}
{\rm e}^{i ( \mbox{\boldmath${\scriptstyle p}$}_1
+ \mbox{\boldmath${\scriptstyle p}$}_2
- \mbox{\boldmath${\scriptstyle p}$}'_1
- \mbox{\boldmath${\scriptstyle p}$}'_2 ) \cdot
\mbox{\boldmath${\scriptstyle y}$}}
\right)
\nonumber\\
&&\!\!\!\!\!
+ \,
i f^{cab} \,
{\rm tr}
\left\{
t^c
u (\mbox{\boldmath$x$})
t^d t^e
u^\dagger (\mbox{\boldmath$y$})
\right\}
\left(
\frac{1}{\bar u \mbox{\boldmath$p$}_1^2}
{\rm e}^{i ( \mbox{\boldmath${\scriptstyle p}$}_1
+ \mbox{\boldmath${\scriptstyle p}$}_2
- \mbox{\boldmath${\scriptstyle p}$}'_1
- \mbox{\boldmath${\scriptstyle p}$}'_2 ) \cdot
\mbox{\boldmath${\scriptstyle y}$}}
-
\frac{1}{u \mbox{\boldmath$p$}_2^2}
{\rm e}^{i ( \mbox{\boldmath${\scriptstyle p}$}_1
+ \mbox{\boldmath${\scriptstyle p}$}_2
- \mbox{\boldmath${\scriptstyle p}$}'_1
- \mbox{\boldmath${\scriptstyle p}$}'_2 ) \cdot
\mbox{\boldmath${\scriptstyle x}$}}
\right)
\Bigg\}
\, . \nonumber
\end{eqnarray}
Finally, the diagrams \ref{nlo} ($e$) result into
\begin{eqnarray}
{\cal N} (\mbox{\boldmath$x$}, \mbox{\boldmath$y$})
\!\!\!&=&\!\!\!
16 \frac{\alpha_s^2}{N_c} \int_{0}^{1} du
\int \frac{d p_-}{p_-}
\int d^2 \mbox{\boldmath$z$} \ d^2 \mbox{\boldmath$z$}'
\
{\rm tr}
\left\{
t^a t^b
u (\mbox{\boldmath$x$})
t^c t^d
u^\dagger (\mbox{\boldmath$y$})
\right\}
\,
u^{ac} (\mbox{\boldmath$z$})
u^{bd} (\mbox{\boldmath$z$}')
\\
&&\times
\int \frac{d^2 \mbox{\boldmath$p$}_1}{(2 \pi)^2}
\frac{d^2 \mbox{\boldmath$p$}'_1}{(2 \pi)^2}
\frac{d^2 \mbox{\boldmath$p$}_2}{(2 \pi)^2}
\frac{d^2 \mbox{\boldmath$p$}'_2}{(2 \pi)^2}
\frac{\mbox{\boldmath$p$}_1 \cdot \mbox{\boldmath$p$}'_1
\
\mbox{\boldmath$p$}_2
\cdot
\mbox{\boldmath$p$}'_2
}{
\left( \bar u \mbox{\boldmath$p$}_1^2 + u \mbox{\boldmath$p$}_2^2 \right)
\left( \bar u {\mbox{\boldmath$p$}'_1}^2 + u {\mbox{\boldmath$p$}'_2}^2 \right)
} \nonumber\\
&\times&\!\!\!
\Bigg\{
{\rm e}^{i ( \mbox{\boldmath${\scriptstyle p}$}_1
- \mbox{\boldmath${\scriptstyle p}$}'_1 )\cdot
( \mbox{\boldmath${\scriptstyle x}$}
- \mbox{\boldmath${\scriptstyle z}$} )
+
i ( \mbox{\boldmath${\scriptstyle p}$}_2
- \mbox{\boldmath${\scriptstyle p}$}'_2 )\cdot
( \mbox{\boldmath${\scriptstyle x}$}
- \mbox{\boldmath${\scriptstyle z}$}' )
}
\frac{1}{ \mbox{\boldmath$p$}_1^2 \ {\mbox{\boldmath$p$}'_2}^2 }
+ \
{\rm e}^{i ( \mbox{\boldmath${\scriptstyle p}$}_1
- \mbox{\boldmath${\scriptstyle p}$}'_1 )\cdot
( \mbox{\boldmath${\scriptstyle y}$}
- \mbox{\boldmath${\scriptstyle z}$} )
+
i ( \mbox{\boldmath${\scriptstyle p}$}_2
- \mbox{\boldmath${\scriptstyle p}$}'_2 )\cdot
( \mbox{\boldmath${\scriptstyle y}$}
- \mbox{\boldmath${\scriptstyle z}$}' )
}
\frac{1}{ \mbox{\boldmath$p$}_2^2 \ {\mbox{\boldmath$p$}'_1}^2 }
\Bigg\}
\, , \nonumber
\end{eqnarray}
and \ref{nlo} ($f$) is
\begin{eqnarray}
{\cal N} (\mbox{\boldmath$x$}, \mbox{\boldmath$y$})
\!\!\!&=&\!\!\!
16 \frac{\alpha_s^2}{N_c} \int_{0}^{1} du
\int \frac{d p_-}{p_-}
\int d^2 \mbox{\boldmath$z$} \ d^2 \mbox{\boldmath$z$}'
\
{\rm tr}
\left\{
t^a
u (\mbox{\boldmath$x$})
t^b t^c
u^\dagger (\mbox{\boldmath$y$})
t^d
\right\}
\,
u^{ab} (\mbox{\boldmath$z$})
u^{dc} (\mbox{\boldmath$z$}')
\\
&\times&\!\!\!
\int \frac{d^2 \mbox{\boldmath$p$}_1}{(2 \pi)^2}
\frac{d^2 \mbox{\boldmath$p$}'_1}{(2 \pi)^2}
\frac{d^2 \mbox{\boldmath$p$}_2}{(2 \pi)^2}
\frac{d^2 \mbox{\boldmath$p$}'_2}{(2 \pi)^2}
{\rm e}^{i ( \mbox{\boldmath${\scriptstyle p}$}_1
- \mbox{\boldmath${\scriptstyle p}$}'_1) \cdot
( \mbox{\boldmath${\scriptstyle x}$}
- \mbox{\boldmath${\scriptstyle z}$} )
+
i ( \mbox{\boldmath${\scriptstyle p}$}_2
- \mbox{\boldmath${\scriptstyle p}$}'_2) \cdot
( \mbox{\boldmath${\scriptstyle y}$}
- \mbox{\boldmath${\scriptstyle z}$}' )
}
\frac{1}{u \bar u} \,
\frac{\mbox{\boldmath$p$}_1 \cdot \mbox{\boldmath$p$}'_1
\
\mbox{\boldmath$p$}_2
\cdot
\mbox{\boldmath$p$}'_2
}{
\mbox{\boldmath$p$}_1^2 \ \mbox{\boldmath$p$}_2^2 \
{\mbox{\boldmath$p$}'_1}^2 \ {\mbox{\boldmath$p$}'_2}^2
}
\, . \nonumber
\end{eqnarray}
The color algebra will be done in section \ref{ColorAlgebra2}.

\subsection{Vertex topology.}

The diagrams having vertex topology are dealt with the same technique.
We refer the reader the Appendix \ref{NonAbelianDiag} for the computation
of the diagrams \ref{nlo} ($g$) and present only the final result here.
Namely, we have
\begin{eqnarray}
\label{NonAbel1}
{\cal N} (\mbox{\boldmath$x$}, \mbox{\boldmath$y$})
\!\!\!&=&\!\!\!
- 8 \frac{\alpha_s^2}{N_c} \int_{0}^{1} du
\int \frac{d p_-}{p_-}
\int d^2 \mbox{\boldmath$z$} \ d^2 \mbox{\boldmath$z$}'
\,
u^{ad} (\mbox{\boldmath$z$})
u^{be} (\mbox{\boldmath$z$}')
\\
&&\times \int \frac{d^2 \mbox{\boldmath$p$}_1}{(2 \pi)^2}
\frac{d^2 \mbox{\boldmath$p$}'_1}{(2 \pi)^2}
\frac{d^2 \mbox{\boldmath$p$}_2}{(2 \pi)^2}
\frac{d^2 \mbox{\boldmath$p$}'_2}{(2 \pi)^2}
\frac{
{\rm e}^{- i ( \mbox{\boldmath${\scriptstyle p}$}_1
- \mbox{\boldmath${\scriptstyle p}$}'_1 ) \cdot
\mbox{\boldmath${\scriptstyle z}$}
- i ( \mbox{\boldmath${\scriptstyle p}$}_2
- \mbox{\boldmath${\scriptstyle p}$}'_2 ) \cdot
\mbox{\boldmath${\scriptstyle z}$}'}
}
{
\left( \mbox{\boldmath$p$}'_1 + \mbox{\boldmath$p$}'_2 \right)^2
\left( \bar u \mbox{\boldmath$p$}_1^2 + u \mbox{\boldmath$p$}_2^2 \right)
\left( \bar u {\mbox{\boldmath$p$}'_1}^2 + u {\mbox{\boldmath$p$}'_2}^2 \right)
}
\nonumber\\
&&\times
\left\{
u \bar u \, \mbox{\boldmath$p$}_1 \cdot \mbox{\boldmath$p$}_2 \
\left( {\mbox{\boldmath$p$}'_1}^2 - {\mbox{\boldmath$p$}'_2}^2 \right)
- 2 \bar u \, \mbox{\boldmath$p$}_1 \cdot \mbox{\boldmath$p$}'_1
\ \mbox{\boldmath$p$}_2 \cdot \left( \mbox{\boldmath$p$}'_1
+ \mbox{\boldmath$p$}'_2 \right)
+ 2 u \, \mbox{\boldmath$p$}_2 \cdot \mbox{\boldmath$p$}'_2
\ \mbox{\boldmath$p$}_1 \cdot \left( \mbox{\boldmath$p$}'_1
+ \mbox{\boldmath$p$}'_2 \right)
\right\}
\nonumber\\
&\times&\!\!\!
\Bigg\{
i f^{cde} \,
{\rm tr}
\left\{
t^a t^b
u (\mbox{\boldmath$x$})
t^c
u^\dagger (\mbox{\boldmath$y$})
\right\}
\left(
\frac{1}{\bar u \mbox{\boldmath$p$}_1^2}
{\rm e}^{i ( \mbox{\boldmath${\scriptstyle p}$}_1
+ \mbox{\boldmath${\scriptstyle p}$}_2 ) \cdot
\mbox{\boldmath${\scriptstyle x}$}
- i ( \mbox{\boldmath${\scriptstyle p}$}'_1
+ \mbox{\boldmath${\scriptstyle p}$}'_2 ) \cdot
\mbox{\boldmath${\scriptstyle y}$}}
-
\frac{1}{u \mbox{\boldmath$p$}_2^2}
{\rm e}^{i ( \mbox{\boldmath${\scriptstyle p}$}_1
+ \mbox{\boldmath${\scriptstyle p}$}_2 ) \cdot
\mbox{\boldmath${\scriptstyle y}$}
- i ( \mbox{\boldmath${\scriptstyle p}$}'_1
+ \mbox{\boldmath${\scriptstyle p}$}'_2 ) \cdot
\mbox{\boldmath${\scriptstyle x}$}}
\right)
\nonumber\\
&&\!\!\!\!\!
+ \,
i f^{cab} \,
{\rm tr}
\left\{
t^c
u (\mbox{\boldmath$x$})
t^d t^e
u^\dagger (\mbox{\boldmath$y$})
\right\}
\left(
\frac{1}{\bar u \mbox{\boldmath$p$}_1^2}
{\rm e}^{i ( \mbox{\boldmath${\scriptstyle p}$}_1
+ \mbox{\boldmath${\scriptstyle p}$}_2 ) \cdot
\mbox{\boldmath${\scriptstyle y}$}
- i ( \mbox{\boldmath${\scriptstyle p}$}'_1
+ \mbox{\boldmath${\scriptstyle p}$}'_2 ) \cdot
\mbox{\boldmath${\scriptstyle x}$}}
-
\frac{1}{u \mbox{\boldmath$p$}_2^2}
{\rm e}^{i ( \mbox{\boldmath${\scriptstyle p}$}_1
+ \mbox{\boldmath${\scriptstyle p}$}_2 ) \cdot
\mbox{\boldmath${\scriptstyle x}$}
- i ( \mbox{\boldmath${\scriptstyle p}$}'_1
+ \mbox{\boldmath${\scriptstyle p}$}'_2 ) \cdot
\mbox{\boldmath${\scriptstyle y}$}}
\right)
\Bigg\}
\, , \nonumber
\end{eqnarray}
for \ref{nlo} ($g$),
\begin{eqnarray}
{\cal N} (\mbox{\boldmath$x$}, \mbox{\boldmath$y$})
\!\!\!&=&\!\!\!
8 \frac{\alpha_s^2}{N_c} \int_{0}^{1} du
\int \frac{d p_-}{p_-}
\int d^2 \mbox{\boldmath$z$} \ d^2 \mbox{\boldmath$z$}'
\,
u^{ad} (\mbox{\boldmath$z$})
u^{be} (\mbox{\boldmath$z$}')
\\
&&\qquad\qquad\qquad\qquad\qquad\times
\Bigg\{
i f^{bca}
{\rm tr}
\left\{
u (\mbox{\boldmath$x$})
t^d t^e
u^\dagger (\mbox{\boldmath$y$})
t^c
\right\}
+
i f^{dce}
{\rm tr}
\left\{
u (\mbox{\boldmath$x$})
t^c
u^\dagger (\mbox{\boldmath$y$})
t^b t^a
\right\}
\Bigg\}
\nonumber\\
&\times&\!\!\!
\int \frac{d^2 \mbox{\boldmath$p$}_1}{(2 \pi)^2}
\frac{d^2 \mbox{\boldmath$p$}'_1}{(2 \pi)^2}
\frac{d^2 \mbox{\boldmath$p$}_2}{(2 \pi)^2}
\frac{d^2 \mbox{\boldmath$p$}'_2}{(2 \pi)^2}
\frac{
{\rm e}^{
- i ( \mbox{\boldmath${\scriptstyle p}$}_1
- \mbox{\boldmath${\scriptstyle p}$}'_1 ) \cdot
\mbox{\boldmath${\scriptstyle z}$}
- i ( \mbox{\boldmath${\scriptstyle p}$}_2
- \mbox{\boldmath${\scriptstyle p}$}'_2 ) \cdot
\mbox{\boldmath${\scriptstyle z}$}' }
}
{
\mbox{\boldmath$p$}_1^2 \mbox{\boldmath$p$}_2^2
\left( \mbox{\boldmath$p$}'_1 + \mbox{\boldmath$p$}'_2 \right)^2
\left( \bar u {\mbox{\boldmath$p$}'_1}^2 + u {\mbox{\boldmath$p$}'_2}^2 \right)
}
\nonumber\\
&&\qquad\qquad\qquad\qquad\qquad\times
\left(
{\rm e}^{i ( \mbox{\boldmath${\scriptstyle p}$}_1
- \mbox{\boldmath${\scriptstyle p}$}'_1 - \mbox{\boldmath${\scriptstyle p}$}'_2 )
\cdot \mbox{\boldmath${\scriptstyle x}$}
+ i \mbox{\boldmath${\scriptstyle p}$}_2 \mbox{\boldmath${\scriptstyle y}$}
}
-
{\rm e}^{i ( \mbox{\boldmath${\scriptstyle p}$}_2
- \mbox{\boldmath${\scriptstyle p}$}'_1 - \mbox{\boldmath${\scriptstyle p}$}'_2 )
\cdot \mbox{\boldmath${\scriptstyle y}$}
+ i \mbox{\boldmath${\scriptstyle p}$}_1 \mbox{\boldmath${\scriptstyle x}$} }
\right)
\nonumber\\
&\times&\!\!\!
\frac{1}{u \bar u}
\left\{
u \bar u \, \mbox{\boldmath$p$}_1 \cdot \mbox{\boldmath$p$}_2 \
\left( {\mbox{\boldmath$p$}'_1}^2 - {\mbox{\boldmath$p$}'_2}^2 \right)
- 2 \bar u \, \mbox{\boldmath$p$}_1 \cdot \mbox{\boldmath$p$}'_1
\ \mbox{\boldmath$p$}_2 \cdot \left( \mbox{\boldmath$p$}'_1
+ \mbox{\boldmath$p$}'_2 \right)
+ 2 u \, \mbox{\boldmath$p$}_2 \cdot \mbox{\boldmath$p$}'_2
\ \mbox{\boldmath$p$}_1 \cdot \left( \mbox{\boldmath$p$}'_1
+ \mbox{\boldmath$p$}'_2 \right)
\right\}
\, , \nonumber
\end{eqnarray}
for \ref{nlo} ($h$), and
\begin{eqnarray}
{\cal N} (\mbox{\boldmath$x$}, \mbox{\boldmath$y$})
\!\!\!&=&\!\!\!
- 16 \frac{\alpha_s^2}{N_c} \int_{0}^{1} du
\int \frac{d p_-}{p_-}
\int d^2 \mbox{\boldmath$z$} \ d^2 \mbox{\boldmath$z$}'
\
{\rm tr}
\left\{
t^a t^b
u (\mbox{\boldmath$x$})
t^c t^d
u^\dagger (\mbox{\boldmath$y$})
\right\}
\,
u^{bd} (\mbox{\boldmath$z$})
u^{ac} (\mbox{\boldmath$z$}')
\\
&\times&\!\!\!
\int \frac{d^2 \mbox{\boldmath$p$}_1}{(2 \pi)^2}
\frac{d^2 \mbox{\boldmath$p$}'_1}{(2 \pi)^2}
\frac{d^2 \mbox{\boldmath$p$}_2}{(2 \pi)^2}
\frac{d^2 \mbox{\boldmath$p$}'_2}{(2 \pi)^2}
\
{\rm e}^{
-
i ( \mbox{\boldmath${\scriptstyle p}$}_1
- \mbox{\boldmath${\scriptstyle p}$}'_1 )
\cdot \mbox{\boldmath${\scriptstyle z}$}
-
i ( \mbox{\boldmath${\scriptstyle p}$}_2
- \mbox{\boldmath${\scriptstyle p}$}'_2 )
\cdot \mbox{\boldmath${\scriptstyle z}$}'
}
\nonumber\\
&&\qquad\qquad\qquad\qquad\quad\times\
\Bigg\{
{\rm e}^{i ( \mbox{\boldmath${\scriptstyle p}$}_1
+ \mbox{\boldmath${\scriptstyle p}$}_2
- \mbox{\boldmath${\scriptstyle p}$}'_2
) \cdot \mbox{\boldmath${\scriptstyle x}$}
-
i \mbox{\boldmath${\scriptstyle p}$}'_1
\cdot \mbox{\boldmath${\scriptstyle y}$}
}
\frac{1}{u} \,
\frac{\mbox{\boldmath$p$}_1 \cdot \mbox{\boldmath$p$}'_1
\
\mbox{\boldmath$p$}_2
\cdot
\mbox{\boldmath$p$}'_2
}{
\mbox{\boldmath$p$}_2^2 \
{\mbox{\boldmath$p$}'_1}^2 \ {\mbox{\boldmath$p$}'_2}^2
\left( \bar u \mbox{\boldmath$p$}_1^2
+ u \mbox{\boldmath$p$}_2^2 \right)}
\nonumber\\
&&\qquad\qquad\qquad\qquad\quad+\
{\rm e}^{
i ( \mbox{\boldmath${\scriptstyle p}$}_1
- \mbox{\boldmath${\scriptstyle p}$}'_1
- \mbox{\boldmath${\scriptstyle p}$}'_2
) \cdot \mbox{\boldmath${\scriptstyle x}$}
+
i \mbox{\boldmath${\scriptstyle p}$}_2
\cdot \mbox{\boldmath${\scriptstyle y}$}
}
\frac{1}{\bar u} \,
\frac{\mbox{\boldmath$p$}_1 \cdot \mbox{\boldmath$p$}'_1
\
\mbox{\boldmath$p$}_2
\cdot
\mbox{\boldmath$p$}'_2
}{
\mbox{\boldmath$p$}_1^2 \
\mbox{\boldmath$p$}_2^2 \ {\mbox{\boldmath$p$}'_1}^2
\left( \bar u {\mbox{\boldmath$p$}'_1}^2
+ u {\mbox{\boldmath$p$}'_2}^2 \right)}
\nonumber\\
&&\qquad\qquad\qquad\qquad\quad+\
{\rm e}^{
i \mbox{\boldmath${\scriptstyle p}$}_1
\cdot \mbox{\boldmath${\scriptstyle x}$}
+
i ( \mbox{\boldmath${\scriptstyle p}$}_2
- \mbox{\boldmath${\scriptstyle p}$}'_1
- \mbox{\boldmath${\scriptstyle p}$}'_2
) \cdot \mbox{\boldmath${\scriptstyle y}$}
}
\frac{1}{u} \,
\frac{\mbox{\boldmath$p$}_1 \cdot \mbox{\boldmath$p$}'_1
\
\mbox{\boldmath$p$}_2
\cdot
\mbox{\boldmath$p$}'_2
}{
\mbox{\boldmath$p$}_1^2 \
\mbox{\boldmath$p$}_2^2 \ {\mbox{\boldmath$p$}'_2}^2
\left( \bar u {\mbox{\boldmath$p$}'_1}^2
+ u {\mbox{\boldmath$p$}'_2}^2 \right)}
\nonumber\\
&&\qquad\qquad\qquad\qquad\quad+\
{\rm e}^{
- i \mbox{\boldmath${\scriptstyle p}$}'_2
\cdot \mbox{\boldmath${\scriptstyle x}$}
+
i ( \mbox{\boldmath${\scriptstyle p}$}_1
+ \mbox{\boldmath${\scriptstyle p}$}_2
- \mbox{\boldmath${\scriptstyle p}$}'_1
) \cdot \mbox{\boldmath${\scriptstyle y}$}
}
\frac{1}{\bar u} \,
\frac{\mbox{\boldmath$p$}_1 \cdot \mbox{\boldmath$p$}'_1
\
\mbox{\boldmath$p$}_2
\cdot
\mbox{\boldmath$p$}'_2
}{
\mbox{\boldmath$p$}_1^2 \
{\mbox{\boldmath$p$}'_1}^2 \ {\mbox{\boldmath$p$}'_2}^2
\left( \bar u \mbox{\boldmath$p$}_1^2
+ u \mbox{\boldmath$p$}_2^2 \right)}
\Bigg\}
\, , \nonumber
\end{eqnarray}
for \ref{nlo} ($i$), respectively. Note that the last contribution is
nonleading in multicolor limit according to Eq.\ (\ref{ColAl3}).

\subsection{Box topology.}

Finally, the calculation of diagrams having the box topology is
straightforward and we get the result for \ref{nlo} ($j$)
\begin{eqnarray}
{\cal N} (\mbox{\boldmath$x$}, \mbox{\boldmath$y$})
\!\!\!&=&\!\!\!
16 \frac{\alpha_s^2}{N_c} \int_{0}^{1} du
\int \frac{d p_-}{p_-}
\int d^2 \mbox{\boldmath$z$} \ d^2 \mbox{\boldmath$z$}'
\
{\rm tr}
\left\{
t^a
u (\mbox{\boldmath$x$})
t^b t^c
u^\dagger (\mbox{\boldmath$y$})
t^d
\right\}
\,
u^{ac} (\mbox{\boldmath$z$})
u^{db} (\mbox{\boldmath$z$}')
\\
&&\times
\int \frac{d^2 \mbox{\boldmath$p$}_1}{(2 \pi)^2}
\frac{d^2 \mbox{\boldmath$p$}'_1}{(2 \pi)^2}
\frac{d^2 \mbox{\boldmath$p$}_2}{(2 \pi)^2}
\frac{d^2 \mbox{\boldmath$p$}'_2}{(2 \pi)^2}
{\rm e}^{i ( \mbox{\boldmath${\scriptstyle p}$}_1
+ \mbox{\boldmath${\scriptstyle p}$}_2) \cdot
\mbox{\boldmath${\scriptstyle x}$}
- i ( \mbox{\boldmath${\scriptstyle p}$}'_1
+ \mbox{\boldmath${\scriptstyle p}$}'_2) \cdot
\mbox{\boldmath${\scriptstyle y}$}
- i ( \mbox{\boldmath${\scriptstyle p}$}_1
- \mbox{\boldmath${\scriptstyle p}$}'_1 ) \cdot
\mbox{\boldmath${\scriptstyle z}$}
- i ( \mbox{\boldmath${\scriptstyle p}$}_2
- \mbox{\boldmath${\scriptstyle p}$}'_2 ) \cdot
\mbox{\boldmath${\scriptstyle z}$}'}
\frac{1}{u \bar u} \,
\frac{\mbox{\boldmath$p$}_1 \cdot \mbox{\boldmath$p$}'_1
\
\mbox{\boldmath$p$}_2
\cdot
\mbox{\boldmath$p$}'_2
}{
\mbox{\boldmath$p$}_1^2 \ \mbox{\boldmath$p$}_2^2 \
{\mbox{\boldmath$p$}'_1}^2 \ {\mbox{\boldmath$p$}'_2}^2
}
\, . \nonumber
\end{eqnarray}
Here we have used the property (\ref{WLproperty}). As seen from the
topology of the graph, this diagram is non-planar as a result it
is supressed in $N_c$ as compared to leading ones, see Eq.\ (\ref{ColAl3}).
For the diagrams of the type in Fig.\ \ref{nlo} ($k$) we have
\begin{eqnarray}
{\cal N} (\mbox{\boldmath$x$}, \mbox{\boldmath$y$})
\!\!\!&=&\!\!\!
16 \frac{\alpha_s^2}{N_c} \int_{0}^{1} du
\int \frac{d p_-}{p_-}
\int d^2 \mbox{\boldmath$z$} \ d^2 \mbox{\boldmath$z$}'
\
{\rm tr}
\left\{
t^a t^b
u (\mbox{\boldmath$x$})
t^c t^d
u^\dagger (\mbox{\boldmath$y$})
\right\}
\nonumber\\
&&\times
\int \frac{d^2 \mbox{\boldmath$p$}_1}{(2 \pi)^2}
\frac{d^2 \mbox{\boldmath$p$}'_1}{(2 \pi)^2}
\frac{d^2 \mbox{\boldmath$p$}_2}{(2 \pi)^2}
\frac{d^2 \mbox{\boldmath$p$}'_2}{(2 \pi)^2}
\frac{
{\rm e}^{i ( \mbox{\boldmath${\scriptstyle p}$}_1
+ \mbox{\boldmath${\scriptstyle p}$}_2) \cdot
\mbox{\boldmath${\scriptstyle x}$}
- i ( \mbox{\boldmath${\scriptstyle p}$}'_1
+ \mbox{\boldmath${\scriptstyle p}$}'_2) \cdot
\mbox{\boldmath${\scriptstyle y}$}
- i ( \mbox{\boldmath${\scriptstyle p}$}_1
- \mbox{\boldmath${\scriptstyle p}$}'_1 ) \cdot
\mbox{\boldmath${\scriptstyle z}$}
- i ( \mbox{\boldmath${\scriptstyle p}$}_2
- \mbox{\boldmath${\scriptstyle p}$}'_2 ) \cdot
\mbox{\boldmath${\scriptstyle z}$}'}
}{
\left( \bar u \mbox{\boldmath$p$}_1^2 + u \mbox{\boldmath$p$}_2^2 \right)
\left( \bar u {\mbox{\boldmath$p$}'_1}^2 + u {\mbox{\boldmath$p$}'_2}^2 \right)
}
\nonumber\\
&\times&\!\!\!
\mbox{\boldmath$p$}_1
\cdot
\mbox{\boldmath$p$}'_1
\
\mbox{\boldmath$p$}_2
\cdot
\mbox{\boldmath$p$}'_2
\Bigg\{
u^{ac} (\mbox{\boldmath$z$})
u^{bd} (\mbox{\boldmath$z$}')
\left(
\frac{u}{\bar u}
\frac{1}{\mbox{\boldmath$p$}_1^2 \ {\mbox{\boldmath$p$}'_1}^2}
+
\frac{\bar u}{u}
\frac{1}{\mbox{\boldmath$p$}_2^2 \ {\mbox{\boldmath$p$}'_2}^2}
\right)
+
u^{ad} (\mbox{\boldmath$z$})
u^{bc} (\mbox{\boldmath$z$}')
\frac{2}{\mbox{\boldmath$p$}_1^2 \ {\mbox{\boldmath$p$}'_2}^2}
\Bigg\} \, .
\end{eqnarray}
The color algebra of the first term in the curly brackets is handled
according to the Eq.\ (\ref{ColAl3}) and is suppressed, while the
second one is reduced via Eq.\ (\ref{ColAl4}). Last but not least,
the diagrams \ref{nlo} ($l$) lead to
\begin{eqnarray}
\label{DiagL}
{\cal N} (\mbox{\boldmath$x$}, \mbox{\boldmath$y$})
\!\!\!&=&\!\!\!
- 32 \frac{\alpha_s^2}{N_c} \int_{0}^{1} du
\int \frac{d p_-}{p_-}
\int d^2 \mbox{\boldmath$z$} \ d^2 \mbox{\boldmath$z$}'
\
{\rm tr}
\left\{
t^a t^b
u (\mbox{\boldmath$x$})
t^c t^d
u^\dagger (\mbox{\boldmath$y$})
\right\}
\,
u^{ad} (\mbox{\boldmath$z$})
u^{bc} (\mbox{\boldmath$z$}')
\\
&\times&\!\!\!
\int \frac{d^2 \mbox{\boldmath$p$}_1}{(2 \pi)^2}
\frac{d^2 \mbox{\boldmath$p$}'_1}{(2 \pi)^2}
\frac{d^2 \mbox{\boldmath$p$}_2}{(2 \pi)^2}
\frac{d^2 \mbox{\boldmath$p$}'_2}{(2 \pi)^2}
{\rm e}^{
-
i ( \mbox{\boldmath${\scriptstyle p}$}_1
- \mbox{\boldmath${\scriptstyle p}$}'_1 )
\cdot \mbox{\boldmath${\scriptstyle z}$}
-
i ( \mbox{\boldmath${\scriptstyle p}$}_2
- \mbox{\boldmath${\scriptstyle p}$}'_2 )
\cdot \mbox{\boldmath${\scriptstyle z}$}'
}
\nonumber\\
&&\qquad\qquad\qquad\qquad\quad\times\
\Bigg\{
{\rm e}^{
i ( \mbox{\boldmath${\scriptstyle p}$}_1
+ \mbox{\boldmath${\scriptstyle p}$}_2
- \mbox{\boldmath${\scriptstyle p}$}'_2
) \cdot \mbox{\boldmath${\scriptstyle x}$}
-
i \mbox{\boldmath${\scriptstyle p}$}'_1
\cdot \mbox{\boldmath${\scriptstyle y}$}
}
\frac{1}{\bar u} \,
\frac{\mbox{\boldmath$p$}_1 \cdot \mbox{\boldmath$p$}'_1
\
\mbox{\boldmath$p$}_2
\cdot
\mbox{\boldmath$p$}'_2
}{
\mbox{\boldmath$p$}_1^2 \
{\mbox{\boldmath$p$}'_1}^2 \ {\mbox{\boldmath$p$}'_2}^2
\left( \bar u \mbox{\boldmath$p$}_1^2 + u \mbox{\boldmath$p$}_2^2 \right)}
\nonumber\\
&&\qquad\qquad\qquad\qquad\quad+\
{\rm e}^{
i ( \mbox{\boldmath${\scriptstyle p}$}_1
- \mbox{\boldmath${\scriptstyle p}$}'_1
- \mbox{\boldmath${\scriptstyle p}$}'_2
) \cdot \mbox{\boldmath${\scriptstyle y}$}
+
i \mbox{\boldmath${\scriptstyle p}$}_2
\cdot \mbox{\boldmath${\scriptstyle x}$}
}
\frac{1}{u} \,
\frac{\mbox{\boldmath$p$}_1 \cdot \mbox{\boldmath$p$}'_1
\
\mbox{\boldmath$p$}_2
\cdot
\mbox{\boldmath$p$}'_2
}{
\mbox{\boldmath$p$}_1^2 \
\mbox{\boldmath$p$}_2^2 \ {\mbox{\boldmath$p$}'_2}^2
\left( \bar u {\mbox{\boldmath$p$}'_1}^2
+ u {\mbox{\boldmath$p$}'_2}^2 \right)}
\Bigg\}
\, . \nonumber
\end{eqnarray}
Having computed all contributions we are ready to discuss their color
properties.

\subsection{Color algebra.}
\label{ColorAlgebra2}

To figure out which contributions are actually suppressed in the large
$N_c$ limit we simply use an obvious equation $t^a = - i \frac{2}{N_c}
f^{abc} t^b t^c$ and the color Fiertz identity $\left( t^a \right){}^i{}_j
\left( t^a \right){}^k{}_l = \frac{1}{2} \delta^i{}_l \delta^k{}_j
- \frac{1}{2 N_c} \delta^i{}_j \delta^k{}_l$ a number of times
until we get rid of all Gell-Mann matrices in between the $u$-matrices.
This procedure leads to the following reduction formula for the structure
of diagrams ($a$) and ($b$)
\begin{eqnarray}
\label{ColAl1}
4 \, {\rm tr}
\left\{
t^a
u (\mbox{\boldmath$x$})
t^d
u^\dagger (\mbox{\boldmath$y$})
\right\}
\, f^{abc} f^{def} \,
u^{be} (\mbox{\boldmath$z$})
u^{cf} (\mbox{\boldmath$z$}')
\!\!\!&=&\!\!\!
{\rm tr}
\left\{
u (\mbox{\boldmath$x$}) u^\dagger (\mbox{\boldmath$z$})
\right\}
{\rm tr}
\left\{
u (\mbox{\boldmath$z$}) u^\dagger (\mbox{\boldmath$z$}')
\right\}
{\rm tr}
\left\{
u (\mbox{\boldmath$z$}') u^\dagger (\mbox{\boldmath$y$})
\right\}
\nonumber\\
&+&\!\!\!
{\rm tr}
\left\{
u (\mbox{\boldmath$x$}) u^\dagger (\mbox{\boldmath$z$}')
\right\}
{\rm tr}
\left\{
u (\mbox{\boldmath$z$}') u^\dagger (\mbox{\boldmath$z$})
\right\}
{\rm tr}
\left\{
u (\mbox{\boldmath$z$}) u^\dagger (\mbox{\boldmath$y$})
\right\}
\nonumber\\
- {\rm tr}
\left\{
u (\mbox{\boldmath$x$}) u^\dagger (\mbox{\boldmath$z$})
u (\mbox{\boldmath$z$}') u^\dagger (\mbox{\boldmath$y$})
u (\mbox{\boldmath$z$}) u^\dagger (\mbox{\boldmath$z$}')
\right\}
\!\!\!&-&\!\!\!
{\rm tr}
\left\{
u (\mbox{\boldmath$x$}) u^\dagger (\mbox{\boldmath$z$}')
u (\mbox{\boldmath$z$}) u^\dagger (\mbox{\boldmath$y$})
u (\mbox{\boldmath$z$}') u^\dagger (\mbox{\boldmath$z$})
\right\}
\, .
\end{eqnarray}
Obviously, the last two terms in this equality are $1/N_c^2$
power suppressed as compared to the first two.

The reduction of other diagrams is accomplished by means of the results,
\begin{eqnarray}
\label{ColAl2}
4 \, i f^{cab} \,
{\rm tr}
\left\{
t^c
u (\mbox{\boldmath$x$})
t^d t^e
u^\dagger (\mbox{\boldmath$y$})
\right\}
u^{ad} (\mbox{\boldmath$z$})
u^{be} (\mbox{\boldmath$z$}')
\!\!\!&=&\!\!\!
{\rm tr}
\left\{
u (\mbox{\boldmath$x$}) u^\dagger (\mbox{\boldmath$z$})
u (\mbox{\boldmath$z$}') u^\dagger (\mbox{\boldmath$y$})
u (\mbox{\boldmath$z$}) u^\dagger (\mbox{\boldmath$z$}')
\right\}
\\
&-&\!\!\!
{\rm tr}
\left\{
u (\mbox{\boldmath$x$}) u^\dagger (\mbox{\boldmath$z$})
\right\}
{\rm tr}
\left\{
u (\mbox{\boldmath$z$}) u^\dagger (\mbox{\boldmath$z$}')
\right\}
{\rm tr}
\left\{
u (\mbox{\boldmath$z$}') u^\dagger (\mbox{\boldmath$y$})
\right\}
\, , \nonumber
\end{eqnarray}
for diagrams ($d$), ($g$) and ($h$),
\begin{eqnarray}
\label{ColAl3}
4 \, {\rm tr}
\left\{
t^a
u (\mbox{\boldmath$x$})
t^b t^c
u^\dagger (\mbox{\boldmath$y$})
t^d
\right\}
\,
u^{ac} (\mbox{\boldmath$z$})
u^{db} (\mbox{\boldmath$z$}')
\!\!\!&=&\!\!\!
{\rm tr}
\left\{
u (\mbox{\boldmath$x$}) u^\dagger (\mbox{\boldmath$z$}')
u (\mbox{\boldmath$z$}) u^\dagger (\mbox{\boldmath$y$})
u (\mbox{\boldmath$z$}') u^\dagger (\mbox{\boldmath$z$})
\right\} \\
-
\frac{1}{N_c}
{\rm tr}
\left\{
u (\mbox{\boldmath$x$}) u^\dagger (\mbox{\boldmath$z$})
\right\}
{\rm tr}
\left\{
u (\mbox{\boldmath$z$}) u^\dagger (\mbox{\boldmath$y$})
\right\}
\!\!\!&-&\!\!\!
\frac{1}{N_c}
{\rm tr}
\left\{
u (\mbox{\boldmath$x$}) u^\dagger (\mbox{\boldmath$z$}')
\right\}
{\rm tr}
\left\{
u (\mbox{\boldmath$z$}') u^\dagger (\mbox{\boldmath$y$})
\right\}
+
\frac{1}{N_c^2}
{\rm tr}
\left\{
u (\mbox{\boldmath$x$}) u^\dagger (\mbox{\boldmath$y$})
\right\}
\, , \nonumber
\end{eqnarray}
for ($e$), ($i$), ($j$) and ($k$), and finally
\begin{eqnarray}
\label{ColAl4}
4 \, {\rm tr}
\left\{
t^a t^b
u (\mbox{\boldmath$x$})
t^c t^d
u^\dagger (\mbox{\boldmath$y$})
\right\}
\,
u^{ad} (\mbox{\boldmath$z$})
u^{bc} (\mbox{\boldmath$z$}')
\!\!\!&=&\!\!\!
{\rm tr}
\left\{
u (\mbox{\boldmath$x$}) u^\dagger (\mbox{\boldmath$z$}')
\right\}
{\rm tr}
\left\{
u (\mbox{\boldmath$z$}') u^\dagger (\mbox{\boldmath$z$})
\right\}
{\rm tr}
\left\{
u (\mbox{\boldmath$z$}) u^\dagger (\mbox{\boldmath$y$})
\right\} \\
-
\frac{1}{N_c}
{\rm tr}
\left\{
u (\mbox{\boldmath$x$}) u^\dagger (\mbox{\boldmath$z$})
\right\}
{\rm tr}
\left\{
u (\mbox{\boldmath$z$}) u^\dagger (\mbox{\boldmath$y$})
\right\}
\!\!\!&-&\!\!\!
\frac{1}{N_c}
{\rm tr}
\left\{
u (\mbox{\boldmath$x$}) u^\dagger (\mbox{\boldmath$z$}')
\right\}
{\rm tr}
\left\{
u (\mbox{\boldmath$z$}') u^\dagger (\mbox{\boldmath$y$})
\right\}
+
\frac{1}{N_c^2}
{\rm tr}
\left\{
u (\mbox{\boldmath$x$}) u^\dagger (\mbox{\boldmath$y$})
\right\}
\, , \nonumber
\end{eqnarray}
for ($c$), ($f$), ($k$) and ($l$).

In our consequent discussion we limit ourselves to the consideration
of the multicolor limit. Therefore, we suppress all $1/N_c$ effects
in the above equations, i.e., we keep the first two terms in Eq.\
(\ref{ColAl1}), the second one in Eq.\ (\ref{ColAl2}), and the first
in (\ref{ColAl4}).

\subsection{Subtraction of multi-Regge kinematics.}

An immediate feature which is transparent in the expressions we have
derived for contributions of particular graphs is the presence of
the double-logarithmic situation, $\left( \int d p_-/p_- \right)
\left( \int d u/u \right)$. Obviously, this is an expected result
and it corresponds to the iteration of the non-linear term of the
leading order equation ${\cal K}_2 {\cal N} \otimes {\cal K}_2 {\cal N}^2
+ {\cal K}_2 {\cal N}^2 \otimes {\cal K}_2 {\cal N}$:
\begin{equation}
{\cal N} (\mbox{\boldmath$x$}, \mbox{\boldmath$y$})
= 2
\left(
\frac{\alpha_s}{2\pi^2} N_c
\ln \frac{1}{x_{\rm B}}
\right)^2
\int d^2 \mbox{\boldmath$z$} \, d^2 \mbox{\boldmath$z$}'
\frac{( \mbox{\boldmath$x$} - \mbox{\boldmath$y$} )^2 }
{( \mbox{\boldmath$x$} - \mbox{\boldmath$z$} )^2
( \mbox{\boldmath$z$} - \mbox{\boldmath$z$}' )^2
( \mbox{\boldmath$z$}' - \mbox{\boldmath$y$} )^2}
{\cal N} (\mbox{\boldmath$x$}, \mbox{\boldmath$z$})
{\cal N} (\mbox{\boldmath$z$}, \mbox{\boldmath$z$}')
{\cal N} (\mbox{\boldmath$z$}', \mbox{\boldmath$y$})
\, .
\end{equation}
It has to be subtracted. The present case is similar to the one encountered
in the computation of next-to-leading logarithmic corrections to the BFKL
kernel \cite{FadKotLip97} and was handled by cutting off the $u$-integral
\begin{equation}
\label{cutoffUintegral}
\int d u \to \int_{\delta}^{1 - \delta} du \, ,
\end{equation}
which corresponds to  a cutoff in the invariant mass of the produced
parton system, and omitting the term ${\rm coeff.} \cdot \ln 1/\delta$.

The double logarithmic part of the diagrams we have just computed is
\begin{equation}
{\cal N} (\mbox{\boldmath$x$}, \mbox{\boldmath$y$})
= - 2
\left( \frac{\alpha_s}{2\pi^2} N_c \right)^2
\ln \frac{1}{x_{\rm B}} \ln \frac{1}{\delta}
\int d^2 \mbox{\boldmath$z$} \, d^2 \mbox{\boldmath$z$}'
\left( \sum_{\alpha} {\cal J}_{(\alpha)} \right)
{\cal N} (\mbox{\boldmath$x$}, \mbox{\boldmath$z$})
{\cal N} (\mbox{\boldmath$z$}, \mbox{\boldmath$z$}')
{\cal N} (\mbox{\boldmath$z$}', \mbox{\boldmath$y$})
\, ,
\end{equation}
with ${\cal J}_{(\alpha)}$ given by the expressions on the
diagram-by-diagram basis
\begin{eqnarray}
{\cal J}_{(a)}
\!\!\!&=&\!\!\!
\frac{1}{( \mbox{\boldmath$z$} - \mbox{\boldmath$z$}' )^2}
\left\{
\frac{
( \mbox{\boldmath$x$} - \mbox{\boldmath$z$}' ) \cdot
( \mbox{\boldmath$y$} - \mbox{\boldmath$z$}' )
}{
( \mbox{\boldmath$x$} - \mbox{\boldmath$z$}' )^2
( \mbox{\boldmath$y$} - \mbox{\boldmath$z$}' )^2}
+
\frac{
( \mbox{\boldmath$x$} - \mbox{\boldmath$z$} ) \cdot
( \mbox{\boldmath$y$} - \mbox{\boldmath$z$} )
}{
( \mbox{\boldmath$x$} - \mbox{\boldmath$z$} )^2
( \mbox{\boldmath$y$} - \mbox{\boldmath$z$} )^2}
\right\}
\, , \\
{\cal J}_{(b)}
\!\!\!&=&\!\!\!
- \frac{1}{2}
\frac{1}{( \mbox{\boldmath$z$}' - \mbox{\boldmath$z$} )^2}
\left\{
\frac{1}{
( \mbox{\boldmath$x$} - \mbox{\boldmath$z$} )^2}
+
\frac{1}{
( \mbox{\boldmath$x$} - \mbox{\boldmath$z$}' )^2}
+
\frac{1}{
( \mbox{\boldmath$y$} - \mbox{\boldmath$z$} )^2}
+
\frac{1}{
( \mbox{\boldmath$y$} - \mbox{\boldmath$z$}' )^2}
\right\}
\, , \nonumber\\
{\cal J}_{(c)}
\!\!\!&=&\!\!\!
- \frac{1}{2}
\frac{1}{
( \mbox{\boldmath$x$} - \mbox{\boldmath$z$} )^2
( \mbox{\boldmath$x$} - \mbox{\boldmath$z$}' )^2}
- \frac{1}{2}
\frac{1}{
( \mbox{\boldmath$y$} - \mbox{\boldmath$z$} )^2
( \mbox{\boldmath$y$} - \mbox{\boldmath$z$}' )^2}
\, , \nonumber\\
{\cal J}_{(d)}
\!\!\!&=&\!\!\!
\frac{1}{( \mbox{\boldmath$y$} - \mbox{\boldmath$z$} )^2}
\frac{
( \mbox{\boldmath$z$}' - \mbox{\boldmath$y$} ) \cdot
( \mbox{\boldmath$z$}' - \mbox{\boldmath$z$} )
}{
( \mbox{\boldmath$z$}' - \mbox{\boldmath$y$} )^2
( \mbox{\boldmath$z$}' - \mbox{\boldmath$z$} )^2}
+
\frac{1}{( \mbox{\boldmath$z$}' - \mbox{\boldmath$x$} )^2}
\frac{
( \mbox{\boldmath$z$} - \mbox{\boldmath$x$} ) \cdot
( \mbox{\boldmath$z$} - \mbox{\boldmath$z$}' )
}{
( \mbox{\boldmath$z$} - \mbox{\boldmath$x$} )^2
( \mbox{\boldmath$z$} - \mbox{\boldmath$z$}' )^2}
\, , \nonumber\\
{\cal J}_{(f)}
\!\!\!&=&\!\!\!
- \frac{1}{
( \mbox{\boldmath$x$} - \mbox{\boldmath$z$} )^2
( \mbox{\boldmath$y$} - \mbox{\boldmath$z$}' )^2}
\, , \nonumber\\
{\cal J}_{(g)}
\!\!\!&=&\!\!\!
- \frac{
( \mbox{\boldmath$x$} - \mbox{\boldmath$z$} ) \cdot
( \mbox{\boldmath$z$}' - \mbox{\boldmath$z$} )
}{
( \mbox{\boldmath$x$} - \mbox{\boldmath$z$} )^2
( \mbox{\boldmath$z$}' - \mbox{\boldmath$z$} )^2}
\frac{
( \mbox{\boldmath$x$} - \mbox{\boldmath$z$}' ) \cdot
( \mbox{\boldmath$y$} - \mbox{\boldmath$z$}' )
}{
( \mbox{\boldmath$x$} - \mbox{\boldmath$z$}' )^2
( \mbox{\boldmath$y$} - \mbox{\boldmath$z$}' )^2}
-
\frac{
( \mbox{\boldmath$x$} - \mbox{\boldmath$z$} ) \cdot
( \mbox{\boldmath$y$} - \mbox{\boldmath$z$} )
}{
( \mbox{\boldmath$x$} - \mbox{\boldmath$z$} )^2
( \mbox{\boldmath$y$} - \mbox{\boldmath$z$} )^2}
\frac{
( \mbox{\boldmath$z$} - \mbox{\boldmath$z$}' ) \cdot
( \mbox{\boldmath$y$} - \mbox{\boldmath$z$}' )
}{
( \mbox{\boldmath$z$} - \mbox{\boldmath$z$}' )^2
( \mbox{\boldmath$y$} - \mbox{\boldmath$z$}' )^2}
\, , \nonumber\\
{\cal J}_{(h)}
\!\!\!&=&\!\!\!
{\cal J}_{(g)}
+
\frac{1}{( \mbox{\boldmath$x$} - \mbox{\boldmath$z$} )^2}
\frac{
( \mbox{\boldmath$z$}' - \mbox{\boldmath$y$} ) \cdot
( \mbox{\boldmath$z$}' - \mbox{\boldmath$z$} )
}{
( \mbox{\boldmath$z$}' - \mbox{\boldmath$y$} )^2
( \mbox{\boldmath$z$}' - \mbox{\boldmath$z$} )^2}
+
\frac{1}{( \mbox{\boldmath$z$}' - \mbox{\boldmath$y$} )^2}
\frac{
( \mbox{\boldmath$z$} - \mbox{\boldmath$x$} ) \cdot
( \mbox{\boldmath$z$} - \mbox{\boldmath$z$}' )
}{
( \mbox{\boldmath$z$} - \mbox{\boldmath$x$} )^2
( \mbox{\boldmath$z$} - \mbox{\boldmath$z$}' )^2}
\, , \nonumber\\
{\cal J}_{(l)}
\!\!\!&=&\!\!\!
\frac{1}{( \mbox{\boldmath$x$} - \mbox{\boldmath$z$} )^2}
\frac{
( \mbox{\boldmath$z$}' - \mbox{\boldmath$x$} ) \cdot
( \mbox{\boldmath$z$}' - \mbox{\boldmath$y$} )
}{
( \mbox{\boldmath$z$}' - \mbox{\boldmath$x$} )^2
( \mbox{\boldmath$z$}' - \mbox{\boldmath$y$} )^2}
+
\frac{1}{( \mbox{\boldmath$z$}' - \mbox{\boldmath$y$} )^2}
\frac{
( \mbox{\boldmath$z$} - \mbox{\boldmath$x$} ) \cdot
( \mbox{\boldmath$z$} - \mbox{\boldmath$y$} )
}{
( \mbox{\boldmath$z$} - \mbox{\boldmath$x$} )^2
( \mbox{\boldmath$z$} - \mbox{\boldmath$y$} )^2}
\, . \nonumber
\end{eqnarray}
In order to compute these Fourier transforms it was enough to use
Eq.\ (\ref{EuclidInt}). Summing up these contributions we obtain
the required result
\begin{equation}
\sum_{\alpha} {\cal J}_{(\alpha)}
= - \frac{( \mbox{\boldmath$x$} - \mbox{\boldmath$y$} )^2 }
{( \mbox{\boldmath$x$} - \mbox{\boldmath$z$} )^2
( \mbox{\boldmath$z$} - \mbox{\boldmath$z$}' )^2
( \mbox{\boldmath$z$}' - \mbox{\boldmath$y$} )^2} \, .
\end{equation}

\subsection{Evolution kernel ${\cal K}_3$.}
\label{EvolutionKernelK3}

After the multi-Regge kinematics being subtracted, the remainder
defines the evolution kernel in the generalized nonlinear equation
of the term trilinear in dipole densities, i.e.\ ${\cal K}_3$.

The afore mentioned subtraction of multi-Regge kinematics corresponds
to the regularization of the singularities in $u$ \`a la $+$-prescription.
Therefore, in order to extract the quasi-multi-Regge region one has to
make the substitution
\begin{eqnarray}
\frac{1}{u} \to \left[ \frac{1}{u} \right]_+
\equiv \frac{1}{u}
- \delta (u) \int_{0}^{1} \frac{d v}{v} \, ,
\qquad
\frac{1}{\bar u} \to \left[ \frac{1}{\bar u} \right]_+
\equiv \frac{1}{\bar u}
- \delta (\bar u) \int_{0}^{1} \frac{d v}{\bar v} \, ,
\end{eqnarray}
in Eqs.\ (\ref{DiagA}-\ref{DiagL}).

We transform the results into the coordinate representation,
see Appendix \ref{FourierAppendix} for details.
Extracting the factor of coupling constant from the kernel
\begin{equation}
{\cal K}_3 (\mbox{\boldmath$x$}, \mbox{\boldmath$z$},
\mbox{\boldmath$z$}', \mbox{\boldmath$y$})
= \left( \frac{\alpha_s}{2 \pi^2} N_c \right)^2
\left\{
K_3
(\mbox{\boldmath$x$}, \mbox{\boldmath$z$},
\mbox{\boldmath$z$}', \mbox{\boldmath$y$})
+
K_3
(\mbox{\boldmath$y$}, \mbox{\boldmath$z$}',
\mbox{\boldmath$z$}, \mbox{\boldmath$x$})
\right\}
\, ,
\end{equation}
with
\begin{eqnarray*}
K_3 = \sum_{\alpha} K_{3 (\alpha)} \, ,
\end{eqnarray*}
and where we have on the diagram-by-diagram basis for large-$N_c$ part,
\begin{eqnarray}
K_{3 (a)} \!\!\!&=&\!\!\!
\frac{1}{2
( \mbox{\boldmath$z$} - \mbox{\boldmath$z$}' )^4
}
\nonumber\\
&+&\!\!\!
\Bigg\{
2
( \mbox{\boldmath$x$} - \mbox{\boldmath$z$} )
\cdot
( \mbox{\boldmath$y$} - \mbox{\boldmath$z$} )
\frac{
( \mbox{\boldmath$x$} - \mbox{\boldmath$z$}' )^2
( \mbox{\boldmath$z$} - \mbox{\boldmath$z$}' )^2
}{
( \mbox{\boldmath$x$} - \mbox{\boldmath$z$} )^2
}
+
2
( \mbox{\boldmath$x$} - \mbox{\boldmath$z$}' )
\cdot
( \mbox{\boldmath$y$} - \mbox{\boldmath$z$}' )
\frac{
( \mbox{\boldmath$x$} - \mbox{\boldmath$z$} )^2
( \mbox{\boldmath$z$} - \mbox{\boldmath$z$}' )^2
}{
( \mbox{\boldmath$x$} - \mbox{\boldmath$z$}' )^2
}
\nonumber\\
&+&\!\!\!
\left(
( \mbox{\boldmath$x$} - \mbox{\boldmath$z$} )
\cdot
( \mbox{\boldmath$x$} - \mbox{\boldmath$z$}' )
\left(
( \mbox{\boldmath$x$} - \mbox{\boldmath$z$} )^2
+
( \mbox{\boldmath$x$} - \mbox{\boldmath$z$}')^2
\right)
-
( \mbox{\boldmath$x$} - \mbox{\boldmath$z$} )^2
( \mbox{\boldmath$x$} - \mbox{\boldmath$z$}' )^2
\right)
\frac{
( \mbox{\boldmath$y$} - \mbox{\boldmath$z$})^2
-
( \mbox{\boldmath$y$} - \mbox{\boldmath$z$}')^2
}{
( \mbox{\boldmath$x$} - \mbox{\boldmath$z$})^2
-
( \mbox{\boldmath$x$} - \mbox{\boldmath$z$}')^2
}
\nonumber\\
&-&\!\!\!
( \mbox{\boldmath$x$} - \mbox{\boldmath$z$}' )^2
( \mbox{\boldmath$y$} - \mbox{\boldmath$z$} )
\cdot
( \mbox{\boldmath$z$}' - \mbox{\boldmath$z$} )
-
( \mbox{\boldmath$x$} - \mbox{\boldmath$z$} )^2
( \mbox{\boldmath$y$} - \mbox{\boldmath$z$}' )
\cdot
( \mbox{\boldmath$z$} - \mbox{\boldmath$z$}' )
+
2 ( \mbox{\boldmath$z$} - \mbox{\boldmath$z$}' )^4
\nonumber\\
&-&\!\!\!
4
( \mbox{\boldmath$x$} - \mbox{\boldmath$z$}' )
\cdot
( \mbox{\boldmath$y$} - \mbox{\boldmath$z$} )
\
( \mbox{\boldmath$x$} - \mbox{\boldmath$z$} )
\cdot
( \mbox{\boldmath$y$} - \mbox{\boldmath$z$}' )
+
4
( \mbox{\boldmath$x$} - \mbox{\boldmath$z$}' )
\cdot
( \mbox{\boldmath$y$} - \mbox{\boldmath$z$}' )
\
( \mbox{\boldmath$x$} - \mbox{\boldmath$z$} )
\cdot
( \mbox{\boldmath$y$} - \mbox{\boldmath$z$} )
\nonumber\\
&-&\!\!\! 2 ( \mbox{\boldmath$z$} - \mbox{\boldmath$z$}' )^2
\left(
( \mbox{\boldmath$x$} - \mbox{\boldmath$z$} )
\cdot
( \mbox{\boldmath$y$} - \mbox{\boldmath$z$} )
+
( \mbox{\boldmath$x$} - \mbox{\boldmath$z$}' )
\cdot
( \mbox{\boldmath$y$} - \mbox{\boldmath$z$}' )
\right)
\Bigg\}
\nonumber\\
&&\times
\frac{\ell (\mbox{\boldmath$x$})}{
( \mbox{\boldmath$z$} - \mbox{\boldmath$z$}' )^4
( ( \mbox{\boldmath$x$} - \mbox{\boldmath$z$} )^2
( \mbox{\boldmath$y$} - \mbox{\boldmath$z$}' )^2
-
( \mbox{\boldmath$x$} - \mbox{\boldmath$z$}' )^2
( \mbox{\boldmath$y$} - \mbox{\boldmath$z$} )^2 )}
\, ,
\nonumber\\
K_{3 (b)} \!\!\!&=&\!\!\!
- \frac{1}{
( \mbox{\boldmath$x$} - \mbox{\boldmath$z$} )^2
( \mbox{\boldmath$z$} - \mbox{\boldmath$z$}' )^4
( \mbox{\boldmath$z$}' - \mbox{\boldmath$x$} )^2
}
\Bigg\{
( \mbox{\boldmath$z$} - \mbox{\boldmath$x$} )^2
( \mbox{\boldmath$z$}' - \mbox{\boldmath$x$} )^2
+
2
\left(
( \mbox{\boldmath$z$} - \mbox{\boldmath$x$} )
\cdot
( \mbox{\boldmath$z$}' - \mbox{\boldmath$x$} )
\right)^2
\nonumber\\
&+&\!\!\!
\left(
( \mbox{\boldmath$z$} - \mbox{\boldmath$x$} )^2
+
( \mbox{\boldmath$z$}' - \mbox{\boldmath$x$} )^2
\right)
\left(
( \mbox{\boldmath$z$} - \mbox{\boldmath$z$}' )^2
-
( \mbox{\boldmath$z$} - \mbox{\boldmath$x$} )
\cdot
( \mbox{\boldmath$z$}' - \mbox{\boldmath$x$} )
\right)
\Bigg\}
\nonumber\\
&+&\!\!\!
\Bigg\{
\frac{3}{2} ( \mbox{\boldmath$z$} - \mbox{\boldmath$z$}' )^2
+
( \mbox{\boldmath$z$} - \mbox{\boldmath$x$} )
\cdot
( \mbox{\boldmath$z$}' - \mbox{\boldmath$x$} )
-
\frac{
( \mbox{\boldmath$z$} - \mbox{\boldmath$z$}' )^2
}{
( \mbox{\boldmath$z$} - \mbox{\boldmath$x$} )^2
( \mbox{\boldmath$z$}' - \mbox{\boldmath$x$} )^2
}
\left(
( \mbox{\boldmath$z$} - \mbox{\boldmath$x$} )^4
+
( \mbox{\boldmath$z$}' - \mbox{\boldmath$x$} )^4
\right)
\Bigg\}
\nonumber\\
&&\times
\frac{\ell (\mbox{\boldmath$x$})}{
( \mbox{\boldmath$z$} - \mbox{\boldmath$z$}' )^4
\left(
( \mbox{\boldmath$z$} - \mbox{\boldmath$x$} )^2
-
( \mbox{\boldmath$z$}' - \mbox{\boldmath$x$} )^2
\right)
}
\, ,
\nonumber\\
K_{3 (c)} \!\!\!&=&\!\!\!
- \frac{1 +
\ell (\mbox{\boldmath$x$})
}{
( \mbox{\boldmath$z$} - \mbox{\boldmath$x$} )^2
( \mbox{\boldmath$z$}' - \mbox{\boldmath$x$} )^2}
\, ,
\nonumber\\
K_{3 (d)} \!\!\!&=&\!\!\!
\frac{
( \mbox{\boldmath$x$} - \mbox{\boldmath$z$} )
\cdot
( \mbox{\boldmath$x$} - \mbox{\boldmath$z$}' )
+
2 \, ( \mbox{\boldmath$z$} - \mbox{\boldmath$z$}' )^2
}{
( \mbox{\boldmath$x$} - \mbox{\boldmath$z$} )^2
( \mbox{\boldmath$z$} - \mbox{\boldmath$z$}' )^2
( \mbox{\boldmath$x$} - \mbox{\boldmath$z$}' )^2
}
\nonumber\\
&+&\!\!\!
\Bigg\{
2 \,
\frac{
( \mbox{\boldmath$x$} - \mbox{\boldmath$z$} )
\cdot
( \mbox{\boldmath$z$}' - \mbox{\boldmath$z$} )
}{
( \mbox{\boldmath$x$} - \mbox{\boldmath$z$}' )^2
}
-
\frac{
( \mbox{\boldmath$x$} - \mbox{\boldmath$z$} )
\cdot
( \mbox{\boldmath$x$} - \mbox{\boldmath$z$}' )
}{
( \mbox{\boldmath$x$} - \mbox{\boldmath$z$} )^2
-
( \mbox{\boldmath$x$} - \mbox{\boldmath$z$}' )^2
}
\Bigg\}
\frac{
\ell (\mbox{\boldmath$x$})
}{
( \mbox{\boldmath$x$} - \mbox{\boldmath$z$} )^2
( \mbox{\boldmath$z$} - \mbox{\boldmath$z$}')^2
}
\, ,
\nonumber\\
K_{3 (g)} \!\!\!&=&\!\!\!
- \Bigg\{
( \mbox{\boldmath$y$} - \mbox{\boldmath$z$} )
\cdot
( \mbox{\boldmath$y$} - \mbox{\boldmath$z$}' )
\frac{
( \mbox{\boldmath$x$} - \mbox{\boldmath$z$}' )^2
}{
( \mbox{\boldmath$y$} - \mbox{\boldmath$z$}' )^2
}
+
( \mbox{\boldmath$x$} - \mbox{\boldmath$z$} )
\cdot
( \mbox{\boldmath$x$} - \mbox{\boldmath$z$}' )
\frac{
( \mbox{\boldmath$y$} - \mbox{\boldmath$z$} )^2
-
( \mbox{\boldmath$y$} - \mbox{\boldmath$z$}' )^2
}{
( \mbox{\boldmath$x$} - \mbox{\boldmath$z$} )^2
-
( \mbox{\boldmath$x$} - \mbox{\boldmath$z$}' )^2
}
\nonumber\\
&+&\!\!\! 2 \,
\frac{
( \mbox{\boldmath$x$} - \mbox{\boldmath$z$} )
\cdot
( \mbox{\boldmath$y$} - \mbox{\boldmath$z$} )
}{
( \mbox{\boldmath$x$} - \mbox{\boldmath$z$} )^2
( \mbox{\boldmath$y$} - \mbox{\boldmath$z$}' )^2
}
\left(
( \mbox{\boldmath$y$} - \mbox{\boldmath$z$}' )^2
( \mbox{\boldmath$z$} - \mbox{\boldmath$z$}' )
\cdot
( \mbox{\boldmath$x$} - \mbox{\boldmath$z$}' )
+
( \mbox{\boldmath$x$} - \mbox{\boldmath$z$}' )^2
( \mbox{\boldmath$z$} - \mbox{\boldmath$z$}' )
\cdot
( \mbox{\boldmath$y$} - \mbox{\boldmath$z$}' )
\right)
\nonumber\\
&+&\!\!\! 2 \,
\frac{
( \mbox{\boldmath$x$} - \mbox{\boldmath$z$}' )
\cdot
( \mbox{\boldmath$y$} - \mbox{\boldmath$z$}' )
}{
( \mbox{\boldmath$x$} - \mbox{\boldmath$z$}' )^2
( \mbox{\boldmath$y$} - \mbox{\boldmath$z$}' )^2
}
\left(
( \mbox{\boldmath$y$} - \mbox{\boldmath$z$}' )^2
( \mbox{\boldmath$z$}' - \mbox{\boldmath$z$} )
\cdot
( \mbox{\boldmath$x$} - \mbox{\boldmath$z$} )
+
( \mbox{\boldmath$x$} - \mbox{\boldmath$z$}' )^2
( \mbox{\boldmath$z$}' - \mbox{\boldmath$z$} )
\cdot
( \mbox{\boldmath$y$} - \mbox{\boldmath$z$} )
\right)
\Bigg\}
\nonumber\\
&&\times
\frac{\ell (\mbox{\boldmath$x$})}{
( \mbox{\boldmath$z$} - \mbox{\boldmath$z$}' )^2
\big(
( \mbox{\boldmath$x$} - \mbox{\boldmath$z$} )^2
( \mbox{\boldmath$y$} - \mbox{\boldmath$z$}' )^2
-
( \mbox{\boldmath$x$} - \mbox{\boldmath$z$}' )^2
( \mbox{\boldmath$y$} - \mbox{\boldmath$z$} )^2
\big)
}
\, ,
\nonumber\\
K_{3 (h)} \!\!\!&=&\!\!\!
-
\Bigg\{
( \mbox{\boldmath$x$} - \mbox{\boldmath$z$}' )
\cdot
( \mbox{\boldmath$y$} - \mbox{\boldmath$z$}' )
+
( \mbox{\boldmath$z$} - \mbox{\boldmath$z$}' )
\cdot
( \mbox{\boldmath$y$} - \mbox{\boldmath$z$}' )
+
2 \, \frac{
( \mbox{\boldmath$x$} - \mbox{\boldmath$z$}' )
\cdot
( \mbox{\boldmath$y$} - \mbox{\boldmath$z$}' )
( \mbox{\boldmath$x$} - \mbox{\boldmath$z$} )
\cdot
( \mbox{\boldmath$z$}' - \mbox{\boldmath$z$} )
}{
( \mbox{\boldmath$x$} - \mbox{\boldmath$z$}' )^2
}
\Bigg\}
\nonumber\\
&&\times
\frac{
\ell (\mbox{\boldmath$x$})
}{
( \mbox{\boldmath$z$} - \mbox{\boldmath$z$}' )^2
( \mbox{\boldmath$x$} - \mbox{\boldmath$z$} )^2
( \mbox{\boldmath$y$} - \mbox{\boldmath$z$}' )^2
}
\, ,
\nonumber\\
K_{3 (k)} \!\!\!&=&\!\!\!
\frac{2}{
( \mbox{\boldmath$z$} - \mbox{\boldmath$x$} )^2
( \mbox{\boldmath$z$}' - \mbox{\boldmath$y$} )^2}
\frac{
( \mbox{\boldmath$z$}' - \mbox{\boldmath$x$} ) \cdot
( \mbox{\boldmath$z$}' - \mbox{\boldmath$y$} ) \
( \mbox{\boldmath$z$} - \mbox{\boldmath$x$} ) \cdot
( \mbox{\boldmath$z$} - \mbox{\boldmath$y$} )
}{
( \mbox{\boldmath$x$} - \mbox{\boldmath$z$} )^2
( \mbox{\boldmath$y$} - \mbox{\boldmath$z$}' )^2
-
( \mbox{\boldmath$x$} - \mbox{\boldmath$z$}' )^2
( \mbox{\boldmath$y$} - \mbox{\boldmath$z$} )^2
}
\ell (\mbox{\boldmath$x$})
\, ,
\nonumber\\
K_{3 (l)} \!\!\!&=&\!\!\!
\frac{2}{( \mbox{\boldmath$x$} - \mbox{\boldmath$z$} )^2}
\frac{
( \mbox{\boldmath$z$}' - \mbox{\boldmath$x$} ) \cdot
( \mbox{\boldmath$z$}' - \mbox{\boldmath$y$} )
}{
( \mbox{\boldmath$z$}' - \mbox{\boldmath$x$} )^2
( \mbox{\boldmath$z$}' - \mbox{\boldmath$y$} )^2
}
\ell (\mbox{\boldmath$x$})
\, , \nonumber
\end{eqnarray}
with
\begin{equation}
\ell (\mbox{\boldmath$x$})
\equiv
\ln
\frac{( \mbox{\boldmath$z$} - \mbox{\boldmath$x$} )^2
}{
( \mbox{\boldmath$z$}' - \mbox{\boldmath$x$} )^2
} \, .
\end{equation}
The sum of these contributions results into the total kernel ${\cal K}_3$. The
log-free piece of which simplifies considerably and reads $- \frac{1}{2} \left(
\frac{\alpha_s}{2 \pi^2} N_c \right)^2 ( \mbox{\boldmath$z$} - \mbox{\boldmath$z$}'
)^{- 4}$. On the other we did not observe an essential simplification for the
logarithmic part. At this point it is timely to say that the total kernel
satisfies an important property. It vanishes in the limit $\mbox{\boldmath$y$}
\to \mbox{\boldmath$x$}$,
\begin{eqnarray*}
{\cal K}_{3} (\mbox{\boldmath$x$},
\mbox{\boldmath$z$}, \mbox{\boldmath$z$}', \mbox{\boldmath$x$}) = 0 \, ,
\end{eqnarray*}
reflecting the unitarity property of the Wilson lines (\ref{Unitarity}).
Namely, as $\mbox{\boldmath$y$} \to \mbox{\boldmath$x$}$: the diagrams ($a$)
cancels with $(b)$, and $(d)$ with $(g)$, respectively. Next, the log-free
term of $(c)$ sums to zero with $(k)$. Finally, the remaining contributions
vanish independently.

\section{Conclusions.}

In the present paper we have developed a formalism which allows to
evaluate successively the nonlinearities in the generalized evolution
equation for the dipole densities. As a demonstration of our machinery
we have calculated the kernel ${\cal K}_3$ which enters with the cubic
nonlinearity in the above equation. Presently, we have not discussed
the question of inclusion of the running of the coupling constant into
our formalism since it runs beyond the scope of this paper and requires
a computation of radiative corrections. This will be done elsewhere.

An obvious continuation of our analysis is to perform a (numerical)
study of the evolution equation keeping the ${\cal K}_3$ contribution
and observe how this affects saturation phenomena.

Other major problems for further research include: (i) A derivation of the
above equation from the Mueller's dipole model by computing the radiative
corrections to the dipole decay kernel. (ii) A computation of $1/N_c$
corrections to our result. (iii) A study of the effects due to delocalization
of the color source from the light cone. Recall that the latter has the
shock-wave form (\ref{ExternalField}) only in the asymptotic limit
$x_{\rm B} \to 0$.

\section*{Acknowledgements}

We would like to thank G.P. Korchemsky, L. McLerran, G. Sterman,
R. Venugopalan for useful discussions. A.B. would like to thank
the Theory Group at Jefferson Lab for the hospitality extended
to him during the final stage of the work. This work  was supported
by the US Department of Energy under contracts DE-AC05-84ER40150 (I.B.)
and DE-FG02-93ER40762 (A.B.).

\appendix

\setcounter{section}{0}
\setcounter{equation}{0}
\renewcommand{\theequation}{\Alph{section}.\arabic{equation}}

\section{Calculation of diagrams with nonabelian vertices.}
\label{NonAbelianDiag}

Let us note first that the three-gluon vertices will not contain the
`$+$'-components of momenta after contraction with gluon propagators
since the latter are orthogonal to $n^\star_\mu$, i.e.,
\begin{eqnarray*}
\frac{\partial}{\partial p_{i+}} {\mit\Gamma}_{\mu\nu\rho} = 0 \, .
\end{eqnarray*}
Since the three-gluon vertices lie on different sides with respect to
the shock wave, we have $z_- > 0$, $z'_- < 0$, for $x_- > 0$ and
$y_- < 0$. Then, as can be seen from the explicit form of Eq.\
(\ref{InteractionPropagator}) the `$-$'-component of the shock-wave
propagators traveling through the external field is positive $p_{i-}
> 0$, $i = 1, 2$. As a result of momentum conservation for the
`$-$'-components in the vertices, the momentum of the free external
propagator $p_{3-} = p_{1-} + p_{2-} > 0$. Therefore, one finds that
the $n^\star_\mu n^\star_\nu$-part of the free propagator, see Eq.\
(\ref{FreePropagator}), does not contribute since the poles in
$p_{3+}$ lie on the same side of the imaginary axis. Namely,
\begin{eqnarray}
&&\int_{0}^{\infty} d x_- \, \int_{0}^{\infty} d z_- \,
\int_{- \infty}^{\infty} d p_{3+} \,
\frac{p_{3+}}{p_3^2 + i 0}
{\rm e}^{- i p_{3+} x_-
+ i \left( p_{3+}
- \mbox{\boldmath${\scriptstyle p}$}_1^2 / (2p_{1-})
- \mbox{\boldmath${\scriptstyle p}$}_2^2 / (2p_{2-})
\right) z_-}
\\
&&\quad= \frac{1}{2p_{3-}}
\int_{- \infty}^{\infty} d p_{3+}
\frac{1}{\left( p_{3+}
- \mbox{\boldmath$p$}_1^2 / (2p_{1-})
- \mbox{\boldmath$p$}_2^2 / (2p_{2-})
+ i 0
\right) \left( p_{3+}
- \mbox{\boldmath$p$}_3^2 / (2p_{3-})
+ i 0
\right)} = 0 \, .
\nonumber
\end{eqnarray}
Same result holds for another free propagator connnecting the loop
to the Wilson line.

Thus, the only nonvanishing contribution is generated by the $d^\perp$
part of the free propagators and we have finally
\begin{eqnarray}
{\cal N} (\mbox{\boldmath$x$}, \mbox{\boldmath$y$})
\!\!\!&=&\!\!\!
- 2 \frac{\alpha_s^2}{N_c} \int_{0}^{1} du \, u \bar u
\int d p_- \, p_-
\int d^2 \mbox{\boldmath$z$} \ d^2 \mbox{\boldmath$z$}'
\
{\rm tr}
\left\{
t^a
u (\mbox{\boldmath$x$})
t^d
u^\dagger (\mbox{\boldmath$y$})
\right\}
\, f^{abc} f^{def} \,
u^{be} (\mbox{\boldmath$z$})
u^{cf} (\mbox{\boldmath$z$}')
\nonumber\\
&&\times
\int \frac{d^2 \mbox{\boldmath$p$}_1}{(2 \pi)^2}
\frac{d^2 \mbox{\boldmath$p$}'_1}{(2 \pi)^2}
\frac{d^2 \mbox{\boldmath$p$}_2}{(2 \pi)^2}
\frac{d^2 \mbox{\boldmath$p$}'_2}{(2 \pi)^2}
{\rm e}^{i ( \mbox{\boldmath${\scriptstyle p}$}_1
+ \mbox{\boldmath${\scriptstyle p}$}_2) \cdot
\mbox{\boldmath${\scriptstyle x}$}
- i ( \mbox{\boldmath${\scriptstyle p}$}'_1
+ \mbox{\boldmath${\scriptstyle p}$}'_2) \cdot
\mbox{\boldmath${\scriptstyle y}$}
- i ( \mbox{\boldmath${\scriptstyle p}$}_1
- \mbox{\boldmath${\scriptstyle p}$}'_1 ) \cdot
\mbox{\boldmath${\scriptstyle z}$}
- i ( \mbox{\boldmath${\scriptstyle p}$}_2
- \mbox{\boldmath${\scriptstyle p}$}'_2 ) \cdot
\mbox{\boldmath${\scriptstyle z}$}'}
\nonumber\\
&&\qquad\qquad\times
\frac{D_1 (p_1, p_2, p'_1, p'_2)}{
\left( \mbox{\boldmath$p$}_1 + \mbox{\boldmath$p$}_2 \right)^2
\left( \mbox{\boldmath$p$}'_1 + \mbox{\boldmath$p$}'_2 \right)^2
\left( \bar u \mbox{\boldmath$p$}_1^2 + u \mbox{\boldmath$p$}_2^2 \right)
\left( \bar u {\mbox{\boldmath$p$}'_1}^2 + u {\mbox{\boldmath$p$}'_2}^2 \right)
}
\, ,
\end{eqnarray}
where we have used the substitution $p_{1-} = u p_-$ and $p_{2-} = (1 - u)
p_-$ in order to reduce two integrals w.r.t.\ $p_{i-}$ by
\begin{equation}
\int_{0}^{\infty} dp_{1-} \int_{0}^{\infty} dp_{1-}
= \int_{0}^{\infty} dp_- p_- \int_{0}^{1} du \, .
\end{equation}
Here
\begin{eqnarray}
D_1 (p_1, p_2, p'_1, p'_2) \!\!\!&=&\!\!\!
{\mit\Gamma}_{\mu\phi\nu} (p_1, - p_1 - p_2, p_2)
{\mit\Gamma}_{\rho\chi\sigma} (p'_1, - p'_1 - p'_2, p'_2)
\nonumber\\
&\times&\!\!\!
d^\perp_{+\phi} \left( p_1 + p_2 \right)
d^\perp_{\mu\lambda} \left( p_1 \right)
d^\perp_{\lambda\rho} \left( p'_1 \right)
d^\perp_{\nu\theta} \left( p_2 \right)
d^\perp_{\theta\sigma} \left( p'_2 \right)
d^\perp_{\chi +} \left( p'_1 + p'_2 \right)
\, ,
\end{eqnarray}
where we have  introduced the notation for the three-gluon vertex
\begin{eqnarray}
{\mit\Gamma}_{\mu\nu\rho} (p_1, p_2, p_3)
\equiv
(p_1 - p_2)_\rho g_{\mu \nu}
+ (p_2 - p_3)_\mu g_{\nu \rho}
+ (p_3 - p_1)_\nu g_{\rho \mu} \, .
\end{eqnarray}
A simple computation leads to the result
\begin{eqnarray}
\label{D1}
p_-^2 \left( u \bar u \right) D_1
\!\!\!&=&\!\!\!
u (1 + u)
\Big\{
\left(
\mbox{\boldmath$p$}_1^2 - \mbox{\boldmath$p$}_2^2
\right)
\
\mbox{\boldmath$p$}'_2
\cdot
\mbox{\boldmath$p$}'
+
\left(
{\mbox{\boldmath$p$}'_1}^2 - {\mbox{\boldmath$p$}'_2}^2
\right)
\
\mbox{\boldmath$p$}_2
\cdot
\mbox{\boldmath$p$}
+
2 \,
\mbox{\boldmath$p$}_2
\cdot
\mbox{\boldmath$p$}'
\
\mbox{\boldmath$p$}'_2
\cdot
\mbox{\boldmath$p$}
\Big\}
\nonumber\\
&+&\!\!\!\bar u (1 + \bar u)
\Big\{
\left(
\mbox{\boldmath$p$}_2^2 - \mbox{\boldmath$p$}_1^2
\right)
\
\mbox{\boldmath$p$}'_1
\cdot
\mbox{\boldmath$p$}'
+
\left(
{\mbox{\boldmath$p$}'_2}^2 - {\mbox{\boldmath$p$}'_1}^2
\right)
\
\mbox{\boldmath$p$}_1
\cdot
\mbox{\boldmath$p$}
+
2 \,
\mbox{\boldmath$p$}_1
\cdot
\mbox{\boldmath$p$}'
\
\mbox{\boldmath$p$}'_1
\cdot
\mbox{\boldmath$p$}
\Big\}
\nonumber\\
&-&\!\!\! 2 (1 + u \bar u)
\Big\{
\mbox{\boldmath$p$}_2
\cdot
\mbox{\boldmath$p$}'
\
\mbox{\boldmath$p$}'_1
\cdot
\mbox{\boldmath$p$}
+
\mbox{\boldmath$p$}_1
\cdot
\mbox{\boldmath$p$}'
\
\mbox{\boldmath$p$}'_2
\cdot
\mbox{\boldmath$p$}
\Big\}
- 2 (1 - u \bar u)
\
\mbox{\boldmath$p$}
\cdot
\mbox{\boldmath$p$}'
\left(
\mbox{\boldmath$p$}_1
\cdot
\mbox{\boldmath$p$}'_2
+
\mbox{\boldmath$p$}_2
\cdot
\mbox{\boldmath$p$}'_1
\right)
\nonumber\\
&&\quad
+ \, \frac{2}{u \bar u}
\
\mbox{\boldmath$p$}
\cdot
\mbox{\boldmath$p$}'
\Big\{
u^2 (1 + u^2)
\mbox{\boldmath$p$}_2 \cdot \mbox{\boldmath$p$}'_2
+
\bar u^2 (1 + \bar u^2)
\mbox{\boldmath$p$}_1 \cdot \mbox{\boldmath$p$}'_1
\Big\}
\, .
\end{eqnarray}
Here and everywhere $\bar u \equiv 1 - u$ and
\begin{equation}
\mbox{\boldmath$p$}
=
\mbox{\boldmath$p$}_1 + \mbox{\boldmath$p$}_2 \, , \qquad
\mbox{\boldmath$p$}'
=
\mbox{\boldmath$p$}'_1 + \mbox{\boldmath$p$}'_2 \, .
\end{equation}

The computation of the diagrams with one nonabelian vertex runs along
the same line. E.g., for Fig.\ \ref{nlo} ($g$) we have
\begin{eqnarray}
{\cal N} (\mbox{\boldmath$x$}, \mbox{\boldmath$y$})
\!\!\!&=&\!\!\!
- 8 \frac{\alpha_s^2}{N_c} \int_{0}^{1} du \, u^2 \bar u
\int d p_- \, p_-^2
\int d^2 \mbox{\boldmath$z$} \ d^2 \mbox{\boldmath$z$}'
\
{\rm tr}
\left\{
t^a t^b
u (\mbox{\boldmath$x$})
t^c
u^\dagger (\mbox{\boldmath$y$})
\right\}
\, i f^{cde} \,
u^{ad} (\mbox{\boldmath$z$})
u^{be} (\mbox{\boldmath$z$}')
\nonumber\\
&&\times
\int \frac{d^2 \mbox{\boldmath$p$}_1}{(2 \pi)^2}
\frac{d^2 \mbox{\boldmath$p$}'_1}{(2 \pi)^2}
\frac{d^2 \mbox{\boldmath$p$}_2}{(2 \pi)^2}
\frac{d^2 \mbox{\boldmath$p$}'_2}{(2 \pi)^2}
{\rm e}^{i ( \mbox{\boldmath${\scriptstyle p}$}_1
+ \mbox{\boldmath${\scriptstyle p}$}_2) \cdot
\mbox{\boldmath${\scriptstyle x}$}
- i ( \mbox{\boldmath${\scriptstyle p}$}'_1
+ \mbox{\boldmath${\scriptstyle p}$}'_2) \cdot
\mbox{\boldmath${\scriptstyle y}$}
- i ( \mbox{\boldmath${\scriptstyle p}$}_1
- \mbox{\boldmath${\scriptstyle p}$}'_1 ) \cdot
\mbox{\boldmath${\scriptstyle z}$}
- i ( \mbox{\boldmath${\scriptstyle p}$}_2
- \mbox{\boldmath${\scriptstyle p}$}'_2 ) \cdot
\mbox{\boldmath${\scriptstyle z}$}'}
\nonumber\\
&&\qquad\qquad\times
\frac{D_2 (p_1, p_2, p'_1, p'_2)}{
\mbox{\boldmath$p$}_1^2
\left( \mbox{\boldmath$p$}'_1 + \mbox{\boldmath$p$}'_2 \right)^2
\left( \bar u \mbox{\boldmath$p$}_1^2 + u \mbox{\boldmath$p$}_2^2 \right)
\left( \bar u {\mbox{\boldmath$p$}'_1}^2 + u {\mbox{\boldmath$p$}'_2}^2 \right)
}
\, ,
\end{eqnarray}
with
\begin{eqnarray}
D_2 (p_1, p_2, p'_1, p'_2) =
{\mit\Gamma}_{\mu\nu\rho} (p'_1, - p'_1 - p'_2, p'_2)
d^\perp_{+\nu} \left( p'_1 + p'_2 \right)
d^\perp_{+\lambda} \left( p_1 \right)
d^\perp_{\lambda\mu} \left( p'_1 \right)
d^\perp_{+\theta} \left( p_2 \right)
d^\perp_{\theta\rho} \left( p'_2 \right)
\, .
\end{eqnarray}
It results into
\begin{equation}
p_-^3 \left( u \bar u \right)^2 D_2
= u \bar u \, \mbox{\boldmath$p$}_1 \cdot \mbox{\boldmath$p$}_2 \
\left( {\mbox{\boldmath$p$}'_1}^2 - {\mbox{\boldmath$p$}'_2}^2 \right)
- 2 \bar u \, \mbox{\boldmath$p$}_1 \cdot \mbox{\boldmath$p$}'_1
\ \mbox{\boldmath$p$}_2 \cdot \left( \mbox{\boldmath$p$}'_1
+ \mbox{\boldmath$p$}'_2 \right)
+ 2 u \, \mbox{\boldmath$p$}_2 \cdot \mbox{\boldmath$p$}'_2
\ \mbox{\boldmath$p$}_1 \cdot \left( \mbox{\boldmath$p$}'_1
+ \mbox{\boldmath$p$}'_2 \right) \, .
\end{equation}
Other three diagrams of the same topology can be easily obtained by
means of symmetry arguments. The final result is given in Eq.\
(\ref{NonAbel1}).

\section{Fourier transformation.}
\label{FourierAppendix}
\setcounter{equation}{0}

In this appendix we give technical details on the Fourier transformation.
To this end we use two simple formulae, the Chisholm representation of the
propagator and the $d$-dimensional Euclidean momentum integral,
respectively,
\begin{eqnarray}
\label{Chisholm}
&&\frac{1}{\mbox{\boldmath$p$}^{2 m}}
= \frac{1}{{\mit\Gamma} (m)}
\int_0^\infty d \alpha \, \alpha^{m - 1}
{\rm e}^{- \alpha \mbox{\boldmath${\scriptstyle p}$}^2}\, ,
\\
\label{ExpFourier}
&&\int \frac{d^d \mbox{\boldmath$p$}}{(2 \pi)^d}
{\rm e}^{i \mbox{\boldmath${\scriptstyle p}$} \cdot
\mbox{\boldmath${\scriptstyle z}$}
- A \mbox{\boldmath${\scriptstyle p}$}^2 }
= \frac{1}{ (4 \pi)^{d/2}} \frac{1}{A^{d/2}}
{\rm e}^{- \frac{\mbox{\boldmath${\scriptstyle z}$}^2}{4 A}} \, ,
\end{eqnarray}
and the formula (\ref{EuclidInt}) as well.

For the sake of definiteness, consider the diagram \ref{nlo} ($g$) which
exhibits all features.
\begin{eqnarray}
\label{Eq3g}
{\cal N} (\mbox{\boldmath$x$}, \mbox{\boldmath$y$})
\!\!\!&=&\!\!\!
- 4 \left( \alpha_s N_c \right)^2 \ln \frac{1}{x_{\rm B}}
\int_{0}^{1} du
\int d^2 \mbox{\boldmath$z$} \ d^2 \mbox{\boldmath$z$}'
{\cal N} (\mbox{\boldmath$x$}, \mbox{\boldmath$z$})
{\cal N} (\mbox{\boldmath$z$}, \mbox{\boldmath$z$}')
{\cal N} (\mbox{\boldmath$z$}', \mbox{\boldmath$y$})
\,\\
&&\times \int \frac{d^2 \mbox{\boldmath$p$}_1}{(2 \pi)^2}
\frac{d^2 \mbox{\boldmath$p$}'_1}{(2 \pi)^2}
\frac{d^2 \mbox{\boldmath$p$}_2}{(2 \pi)^2}
\frac{d^2 \mbox{\boldmath$p$}'_2}{(2 \pi)^2}
\frac{
\mbox{\boldmath$p$}_{1 \alpha} \mbox{\boldmath$p$}_{2 \beta}
f_{\alpha \beta} (\mbox{\boldmath$p$}'_1, \mbox{\boldmath$p$}'_2)
}
{
\left( \mbox{\boldmath$p$}'_1 + \mbox{\boldmath$p$}'_2 \right)^2
\left( \bar u \mbox{\boldmath$p$}_1^2 + u \mbox{\boldmath$p$}_2^2 \right)
\left( \bar u {\mbox{\boldmath$p$}'_1}^2 + u {\mbox{\boldmath$p$}'_2}^2 \right)
}
\nonumber\\
&&\times
{\rm e}^{- i ( \mbox{\boldmath${\scriptstyle p}$}_1
- \mbox{\boldmath${\scriptstyle p}$}'_1 ) \cdot
\mbox{\boldmath${\scriptstyle z}$}
- i ( \mbox{\boldmath${\scriptstyle p}$}_2
- \mbox{\boldmath${\scriptstyle p}$}'_2 ) \cdot
\mbox{\boldmath${\scriptstyle z}$}'}
\Bigg\{
\frac{1}{u \mbox{\boldmath$p$}_2^2}
{\rm e}^{i ( \mbox{\boldmath${\scriptstyle p}$}_1
+ \mbox{\boldmath${\scriptstyle p}$}_2 ) \cdot
\mbox{\boldmath${\scriptstyle x}$}
- i ( \mbox{\boldmath${\scriptstyle p}$}'_1
+ \mbox{\boldmath${\scriptstyle p}$}'_2 ) \cdot
\mbox{\boldmath${\scriptstyle y}$}}
-
\frac{1}{\bar u \mbox{\boldmath$p$}_1^2}
{\rm e}^{i ( \mbox{\boldmath${\scriptstyle p}$}_1
+ \mbox{\boldmath${\scriptstyle p}$}_2 ) \cdot
\mbox{\boldmath${\scriptstyle y}$}
- i ( \mbox{\boldmath${\scriptstyle p}$}'_1
+ \mbox{\boldmath${\scriptstyle p}$}'_2 ) \cdot
\mbox{\boldmath${\scriptstyle x}$}}
\Bigg\}
\, . \nonumber
\end{eqnarray}
with
\begin{equation}
\label{Lorentzf}
f_{\alpha \beta} (\mbox{\boldmath$p$}'_1, \mbox{\boldmath$p$}'_2)
=
u \bar u \, \delta_{\alpha\beta}
\left( {\mbox{\boldmath$p$}'_1}^2 - {\mbox{\boldmath$p$}'_2}^2 \right)
- 2 \bar u \, \mbox{\boldmath$p$}'_{1 \alpha}
\left( \mbox{\boldmath$p$}'_1 + \mbox{\boldmath$p$}'_2 \right)_\beta
+ 2 u \, \mbox{\boldmath$p$}'_{2 \beta}
\left( \mbox{\boldmath$p$}'_1 + \mbox{\boldmath$p$}'_2 \right)_\alpha \, .
\end{equation}
The first letters of the Greek alphabet stand for 2D transverse space
with metric $\delta_{\alpha\beta} = - g^\perp_{\alpha\beta} =
{\rm diag} (1, 1)$, $\alpha, \beta, \gamma, ... = 1, 2$. The tensor
structure factorizes and can be evaluated separately. Consider the
first term in curly brackets. The integral over unprimed momenta gives,
using Eqs.\ (\ref{Chisholm},\ref{ExpFourier})
\begin{eqnarray}
&&\int \frac{d^2 \mbox{\boldmath$p$}_1}{(2 \pi)^2}
\frac{d^2 \mbox{\boldmath$p$}_2}{(2 \pi)^2}
{\rm e}^{i \mbox{\boldmath${\scriptstyle p}$}_1
\cdot (
\mbox{\boldmath${\scriptstyle x}$} - \mbox{\boldmath${\scriptstyle z}$} )
+ i \mbox{\boldmath${\scriptstyle p}$}_2
\cdot (
\mbox{\boldmath${\scriptstyle x}$}
-
\mbox{\boldmath${\scriptstyle z}$}' )}
\frac{
\mbox{\boldmath$p$}_{1 \alpha} \mbox{\boldmath$p$}_{2 \beta}
}
{
\mbox{\boldmath$p$}_2^2
\left( \bar u \mbox{\boldmath$p$}_1^2 + u \mbox{\boldmath$p$}_2^2 \right)
}
\\
&&\qquad\qquad\qquad\qquad\qquad\qquad\qquad\qquad =
\left( \frac{i}{2 \pi} \right)^2
\frac{
( \mbox{\boldmath$x$} -  \mbox{\boldmath$z$} )_{\alpha}
( \mbox{\boldmath$x$} -  \mbox{\boldmath$z$}' )_{\beta}
}{
\left( \mbox{\boldmath$x$} - \mbox{\boldmath$z$} \right)^2
\left( \bar u ( \mbox{\boldmath$x$} - \mbox{\boldmath$z$}' )^2
+ u ( \mbox{\boldmath$x$} - \mbox{\boldmath$z$} )^2 \right)
}
\, . \nonumber
\end{eqnarray}
Note that this formulae is correct only for $u < 1$. However, this should not
bother us since the boundary is not reached due to limits on the final state
mass resulting into the cutoff (\ref{cutoffUintegral}).

The primed momenta are integrated out with the formula
\begin{eqnarray}
&&\int \frac{d^2 \mbox{\boldmath$p$}'_1}{(2 \pi)^2}
\frac{d^2 \mbox{\boldmath$p$}'_2}{(2 \pi)^2}
{\rm e}^{ - i \mbox{\boldmath${\scriptstyle p}$}'_1
\cdot ( \mbox{\boldmath${\scriptstyle y}$}
- \mbox{\boldmath${\scriptstyle z}$} )
- i \mbox{\boldmath${\scriptstyle p}$}'_2
\cdot ( \mbox{\boldmath${\scriptstyle y}$}
- \mbox{\boldmath${\scriptstyle z}$}' )
}
\frac{f_{\alpha \beta} (\mbox{\boldmath$p$}'_1 , \mbox{\boldmath$p$}'_2)}{
\left( \mbox{\boldmath$p$}'_1 + \mbox{\boldmath$p$}'_2 \right)^2
\left( \bar u {\mbox{\boldmath$p$}'_1}^2 + u {\mbox{\boldmath$p$}'_2}^2 \right)}
\\
&&\qquad\qquad = \frac{1}{(4 \pi)^2}
f_{\alpha \beta} \left(
- i \frac{\partial}{\partial \mbox{\boldmath$z$}} ,
- i \frac{\partial}{\partial \mbox{\boldmath$z$}'}
\right)
\int_0^\infty \frac{d \rho}{\rho} \int_0^1 \frac{dv}{v}
\exp \left\{ - \frac{\rho}{v}
\left(
u \bar u (\mbox{\boldmath$z$} - \mbox{\boldmath$z$}' )^2
+ v \, \mbox{\boldmath$V$}^2 (\mbox{\boldmath$y$})
\right)
\right\} \, , \nonumber
\end{eqnarray}
where
\begin{equation}
\mbox{\boldmath$V$} (\mbox{\boldmath$y$})
=
u ( \mbox{\boldmath$z$} - \mbox{\boldmath$y$} )
+
\bar u (\mbox{\boldmath$z$}' - \mbox{\boldmath$y$} ) \, .
\end{equation}

The action of differential operators in $f_{\alpha \beta}$ reduces to
substitutions
\begin{eqnarray}
\frac{\partial^2}{\partial\mbox{\boldmath$z$}^2}
-
\frac{\partial^2}{\partial\mbox{\boldmath$z$}^{\prime 2}}
\!\!\!&=&\!\!\!
- 4 \rho
\left\{
2 u - 1
- \frac{\rho}{v} \,
\bigg(
v \, (2 u - 1)
\mbox{\boldmath$V$} (\mbox{\boldmath$y$})
+
2 u \bar u \,
(\mbox{\boldmath$z$} - \mbox{\boldmath$z$}')
\bigg)
\cdot
\mbox{\boldmath$V$} (\mbox{\boldmath$y$})
\right\}
\, ,
\nonumber\\
\frac{\partial}{\partial\mbox{\boldmath$z$}_\alpha}
\left(
\frac{\partial}{\partial\mbox{\boldmath$z$}_\beta}
-
\frac{\partial}{\partial\mbox{\boldmath$z$}'_\beta}
\right)
\!\!\!&=&\!\!\!
- 2 u \rho
\left\{
\delta_{\alpha\beta}
-
2 \frac{\rho}{v} \,
\bigg(
v \mbox{\boldmath$V$}_{\!\!\alpha} (\mbox{\boldmath$y$})
+
\bar u \, (\mbox{\boldmath$z$} - \mbox{\boldmath$z$}')_\alpha
\bigg)
\mbox{\boldmath$V$}_{\!\!\beta} (\mbox{\boldmath$y$})
\right\}
\, ,
\nonumber\\
\frac{\partial}{\partial\mbox{\boldmath$z$}'_\beta}
\left(
\frac{\partial}{\partial\mbox{\boldmath$z$}_\alpha}
-
\frac{\partial}{\partial\mbox{\boldmath$z$}'_\alpha}
\right)
\!\!\!&=&\!\!\!
- 2 \bar u \rho
\left\{
\delta_{\alpha\beta}
-
2 \frac{\rho}{v} \,
\bigg(
v \mbox{\boldmath$V$}_{\!\!\beta} (\mbox{\boldmath$y$})
-
u \, (\mbox{\boldmath$z$} - \mbox{\boldmath$z$}')_\beta
\bigg)
\mbox{\boldmath$V$}_{\!\!\alpha} (\mbox{\boldmath$y$})
\right\}
\, .
\end{eqnarray}
Finally, we get
\begin{eqnarray}
\label{Fourierf}
&&\int \frac{d^2 \mbox{\boldmath$p$}'_1}{(2 \pi)^2}
\frac{d^2 \mbox{\boldmath$p$}'_2}{(2 \pi)^2}
{\rm e}^{ - i \mbox{\boldmath${\scriptstyle p}$}'_1
\cdot ( \mbox{\boldmath${\scriptstyle y}$}
- \mbox{\boldmath${\scriptstyle z}$} )
- i \mbox{\boldmath${\scriptstyle p}$}'_2
\cdot ( \mbox{\boldmath${\scriptstyle y}$}
- \mbox{\boldmath${\scriptstyle z}$}' )
}
\frac{f_{\alpha \beta} (\mbox{\boldmath$p$}'_1 , \mbox{\boldmath$p$}'_2)}{
\left( \mbox{\boldmath$p$}'_1 + \mbox{\boldmath$p$}'_2 \right)^2
\left( \bar u {\mbox{\boldmath$p$}'_1}^2 + u {\mbox{\boldmath$p$}'_2}^2 \right)}
\\
&&\qquad\qquad
= \left( - \frac{i}{2 \pi} \right)^2
\frac{1}{
\left( \mbox{\boldmath$z$} - \mbox{\boldmath$z$}' \right)^2
\left(
u ( \mbox{\boldmath$z$} - \mbox{\boldmath$y$} )^2
+
\bar u ( \mbox{\boldmath$z$}' - \mbox{\boldmath$y$} )^2
\right)
}
\nonumber\\
&&
\times\left\{
u \bar u
(\mbox{\boldmath$z$} - \mbox{\boldmath$z$}')
\cdot
\bigg(
2 \mbox{\boldmath$V$} (\mbox{\boldmath$y$})
+
(1 - 2 u)
(\mbox{\boldmath$z$} - \mbox{\boldmath$z$}')
\bigg)
\delta_{\alpha \beta}
-
2 \bar u
\mbox{\boldmath$V$}_{\!\!\beta} (\mbox{\boldmath$y$})
\,
(\mbox{\boldmath$z$} - \mbox{\boldmath$z$}')_\alpha
-
2 u
\mbox{\boldmath$V$}_{\!\!\alpha} (\mbox{\boldmath$y$})
\,
(\mbox{\boldmath$z$} - \mbox{\boldmath$z$}')_\beta
\right\} \, .
\nonumber
\end{eqnarray}
Note that $2 \mbox{\boldmath$V$} (\mbox{\boldmath$y$})
+ (1 - 2 u) (\mbox{\boldmath$z$} - \mbox{\boldmath$z$}')
= (\mbox{\boldmath$z$}' - \mbox{\boldmath$y$})
+ (\mbox{\boldmath$z$} - \mbox{\boldmath$y$})$ is $u$-independent.
The second term in the curly brackets in (\ref{Eq3g}) is analyzed
along the same line.

Let us note that the calculation of the diagram \ref{nlo} ($a$) does
not present a difficulty either since $D_1$ from Eq.\ (\ref{D1}) is
a `square' of the Lorentz structures $f_{\alpha\beta}$ (\ref{Lorentzf}).
Namely,
\begin{equation}
p_-^2 \left( u \bar u \right) D_1
= \frac{1}{u \bar u}
f_{\alpha \beta} (\mbox{\boldmath$p$}_1, \mbox{\boldmath$p$}_2)
f_{\alpha \beta} (\mbox{\boldmath$p$}'_1, \mbox{\boldmath$p$}'_2) \, .
\end{equation}
Thus, all the Fourier transforms factorize and are given by Eq.\
(\ref{Fourierf}). In the computation of the diagram \ref{nlo} ($a$)
it is instructive to use the identity
\begin{equation}
\mbox{\boldmath$V$} (\mbox{\boldmath$x$})
\cdot
\mbox{\boldmath$V$} (\mbox{\boldmath$y$})
=
\frac{1}{2}
\left\{
\mbox{\boldmath$V$}^2 (\mbox{\boldmath$x$})
+
\mbox{\boldmath$V$}^2 (\mbox{\boldmath$y$})
-
(\mbox{\boldmath$x$} - \mbox{\boldmath$y$})^2
\right\} \, ,
\end{equation}
in order to simplify the algebra.

\end{document}